\documentstyle[12pt,aasms4]{article}
\newcommand{\tc}{\mbox{$\tau_V(0)$}}
\newcommand{\starred}[1]{#1^\star}
\newcommand{\etal}{{\it et~al.\ }}
\lefthead{Bianchi, Ferrara \& Giovanardi}
\righthead{Extinction and Polarization in Galaxies}

\begin{document}

\title{Monte Carlo simulations of dusty spiral galaxies:\\
extinction and polarization properties}

\author{ Simone Bianchi}
\affil{ Univ. di Firenze Dipartimento di Astronomia e Scienza dello Spazio\\
Largo E. Fermi, 5, 50125 Firenze, Italy}
\authoraddr{ Largo E. Fermi, 5, 50125 Firenze, Italy}
\authoremail{sbianchi@arcetri.astro.it}

\author{ Andrea Ferrara and Carlo Giovanardi}
\affil{ Osservatorio Astrofisico di Arcetri\\
Largo E. Fermi, 5, 50125 Firenze, Italy}

\begin{abstract}
We present Monte Carlo simulations of dusty spiral galaxies,
modelled as bulge + disk systems, aimed to study their extinction
and polarization properties. The extinction parameters (absorption and
scattering) of dust grains are calculated from
Mie's theory for a full distribution of sizes and materials; the
radiation transfer is carried on for the four Stokes parameters.
Photometric and polarimetric maps of galaxies of different optical
depths, inclinations and bulge-to-total ratios have been produced
in the B and I bandpasses. As expected, the effect of scattering is
to reduce substantially the extinction for a given
optical depth, in particular for what concerns the obscuration of
bright bulge cores. For the same reason,
scattering reduces also the reddening, as evaluated from B-I maps.
On the other hand the bluing directly due to forward scattering is
hardly appreciable. Radial color gradients are often found.
A comparison with ``sandwich'' models shows that they fail dramatically
to reproduce the extinction - optical depth relation.
The degree of linear polarization produced by scattering is usually of the
order
of a few percent; it increases
with optical depth, and with inclination ($i \le 80^o$). The
polarization pattern is always perpendicular to the major axis, unless
the dust distribution is drastically modified.
There is little local correlation between extinction and polarization
degree and there is a trend of increasing polarization
from the B to the I band.
We discuss implications and relevance of the results for studies
of the structure and morphology of spiral galaxies and of their
interstellar medium.
\end{abstract}



\section{Introduction }

For a long time, since Holmberg's first analysis (Holmberg \markcite{ho58}
1958), disk galaxies, including the gas-rich  late morphological types,
have been taken as largely transparent systems.
Although with various refinements,
this point of view has been shared by the authors of catalogs of galaxies,
such as the relatively recent Revised Shapley-Ames Catalog of Bright
Galaxies (Sandage \& Tamman \markcite{sa81} 1981), and thereby adopted
by the majority of the astronomical community.
Holmberg inferred average values of the internal extinction on
statistical grounds, that is by interpreting the correlation between
inclination and mean surface brightness in a rather limited
sample of 119 spirals of various morphologies.
On the other hand, refined studies of the dust content and distribution
of individual, representative galaxies are scarce. In the edge-on Sc galaxy
NGC 891, Kylafis \& Bahcall \markcite{ky87} (1987) have been able to
determine an overall transparency of the disk, when seen face-on,
in a bandpass around 6000\AA.

Recently, however, some studies have shaken the widespread belief in
transparent disks (Disney, Davies \& Phillipps \markcite{di89}
1989; Valentijn \markcite{va90} 1990) and
the issue has rapidly become one of the most contended  in the
literature on normal galaxies, sometimes with opposite conclusions drawn
by the analysis of the same, or similar, set of data (Valentijn
\markcite{va94} 1994).
The idea of opaque galaxies has gained in the last years a certain
acceptance and has been shared by the recent Third Reference Catalogue
of Bright Galaxies (de Vaucouleurs et al. \markcite{dv91} 1991).

This debate on the optical thickness of galactic disks is relevant to a
number of issues important for extragalactic astronomy. We mention: (i)
the morphology and internal structure of galactic components,
such as bulges, disks, lenses, bars, etc; (ii) the
dark matter content of galaxies and clusters, and (iii) the luminosity
corrections affecting redshift-independent measures of distances.
The last point, in particular, touches on the field of the large scale
structure of the universe, and of the deviations from the Hubble flow.

For these reasons, observations and data analyses are becoming more
refined with particular reference to selection biases (Burstein, Haynes
\& Faber \markcite{bu91} 1991), to redshift information (Cho\mbox{\l}oniewski
\markcite{ch91} 1991; Giovanelli \markcite{gi94} et al. 1994) and to
multiwavelength imaging (Peletier \markcite{pe94} et al.  1994; Evans
\markcite{ev94} 1994).
At the same time the modelling has also become increasingly sophisticated.
In fact, as it has been argued by most of the already cited authors, a
fundamental problem in statistical evaluations of the extinction lies in
the absence of accurate and sensitive observable indicators. Surface
brightness, diameters and colors have often been interpreted in the
framework of models so naive and unrealistic to yield too simplistic and
sometimes wrong results.

A realistic modelling of dusty disks is thus strongly required. Besides,
if we postulate the possibility of large optical thicknesses and
sizeable grain albedos, as it is the case for wavelengths shorter than 3$\mu
$m, then the effect of multiple scattering in composite (bulge + disk)
systems, with dust interspersed with stars, must be included
(e.g. Block \markcite{bl94} et al. 1994).
A remarkable progress on this track is achieved in the
recent paper by Byun, Freeman \& Kylafis \markcite{bfk94}
(1994, hereafter BFK) who have
produced simulations of disk galaxies of various morphology and optical
thickness. Their treatment of the radiative transfer includes exactly
the contribution of the first scattering, and approximately the higher
scattering orders. They adopt a Henyey-Greenstein phase function and a single
dust grain with average optical properties according to Draine \& Lee
\markcite{dl84} (1984).

Scattering events are also expected to introduce polarization in the
otherwise unpolarized emission from normal stars.
Polarimetric studies of normal spirals, both observational and
theoretical ones, are scarce. In view of the current interest in the
properties of dusty disks such scarcity is remarkable since
most of the information about the physical properties of dust grains is
locked into their polarization effects.
Among the few observations of this kind, we mention the detailed studies
of M31, M82, M104, and of the Magellanic clouds; they are
summarized and referenced by Wielebinski \& Krause \markcite{wi93}
(1993).

One of the fields where the analysis and simulation of polarization
maps has been most effective in recent years is star formation, and,
in particular, the study of protostellar envelopes (see Fischer,
Henning \& Yorke, \markcite{fi94}
1994 and references therein). Here the more
recent and detailed models are based on Monte Carlo simulations which
allow for arbitrary distributions of polydispersive mixtures of
spherical dust particles. In all these models any observed polarization
is imputed to transfer effects on light which is otherwise emitted
unpolarized by stars. In particular, since dust grains are usually
assumed to be
spherical, homogeneous and optically isotropic, only the polarizing
effect of scattering is taken into account. It is widely known, anyway,
that in our Galaxy the
linear polarization introduced by the light propagation through the
interstellar medium is best described as the effect of dichroic
absorption by aligned, elongated grains (Spitzer \markcite{sp78}
1978; Gledhill \& Scarrot \markcite{ge89} 1989).
Sometimes, this polarization mechanism has been questioned but, indeed,
only for particular situations, where the presence of a suitable
toroidal magnetic field, that is the main ingredient for the alignment,
is dubious or improbable (Menard \markcite{me89} 1989).

We present here the results of Monte Carlo simulations of realistic
model galaxies within the frame of Mie's scattering theory, that is for
spherical dust grains with homogeneus and isotropic optical properties.
In our study the dust grain is not a single, average one but instead we
allow for a mixture of different materials and a continuous distribution
of size. The phase function is consistently computed from Mie's theory for
each grain.
The radiation transfer is exact to any scattering order and complete for
the four Stokes parameters, so that the output can be analyzed in the
form of polarimetric, as well as photometric, maps.
Stars and dust three-dimensional distributions are the ones commonly
accepted, and similar to those adopted by BFK.
The present paper deals only with scattering events on spherical
grains and does not include any dichroism in the absorption itself. The
polarization properties may, therefore, be different from those
produced by needles rotating, as widely believed,
in a galactic meridian plane.

Although we deal in this study with the simplest of the realistic
cases, that is,  smooth distributions and spherical grains, the
Monte Carlo method is ideal to probe more hostile situations.
The effect of clumping in the obscuring dust, of non-spherical dust
grains and of extinction by gas at short wavelengths are deferred to
future papers.


\section{A dusty galaxy model}
In the following we describe the model distribution here adopted to
simulate the spatial distribution of the photon emitters (stars) and of
the particles contributing to extinction (dust).

\subsection{Stellar distributions}
\label{distri}
We describe the stellar spatial distribution in a spiral galaxy by
two components: a spheroidal bulge and a 3-dimensional disk.
We neglect dark massive halos, since we are only interested in luminous
components, and also possible small-scale inhomogeneities such as spiral
arms and star-forming regions.

\subsubsection{Bulge}
The surface brightness profiles of spheroidal systems is usually well
described by the $R^{1/4}$ law (de Vaucouleurs \markcite{dv59} 1959):
\begin{equation}
I(R) = I_e ~10^{-3.33 \left[(R/R_e)^{1/4}-1\right]},
\label{R14}
\end{equation}

\noindent where $R_e$ is the equivalent radius, and $I_e$ is the surface
brightness of the corresponding isophote; both $R$ and $R_e$ are
measured on the sky plane.
For numerical purpose, it is convenient to have an analytical
expression for the bulge luminosity density $\rho^{(b)}$, which can be
obtained deprojecting the profiles. Since there is no simple, analytical
expression for $\rho^{(b)}$ corresponding to equation (\ref{R14})
(see Young \markcite{yo76} 1976) we choose to adopt the Jaffe
distribution which corresponds to the
luminosity density (Jaffe \markcite{ja83} 1983):
\begin{equation}
\rho^{(b)}(\tilde{r}) = {\rho^{(b)}_{0}\over
(\tilde{r}/r_b)^2 \left[1+ (\tilde{r}/r_b)\right]^2},
\label{jaffe}
\end{equation}

\noindent where $\tilde{r}$ is the physical distance from the galactic center,
and $r_b$ is the scale radius for wich we adopt $r_b = 1.16\;R_e$ (see
Appendix~\ref{asas}).

In principle, there is no need for an external truncation radius for
the Jaffe distribution since the volume integral converges for
$\tilde{r} \rightarrow \infty$; for numerical purposes, though, we
have introduced a truncation radius  $r^{(b)}_{max}=n r_b$.
In fact, in absence of an external truncation, at large radii the bulge's
light will dominate once again that of the (exponential) disk.
In the simulations presented here, we have adopted the values
$R_e=1.60$~kpc, $n= 8$. The value for $R_e$ is close to the one
inferred for our Galaxy; the value for $n$ is such that $r^{(b)}_{max}$
corresponds to the point where the ratio between the bulge luminosity
density and the density of the adopted disk is maximum (Sect.~\ref{dischi}).

\subsubsection{Disk}
\label{dischi}
The assumed luminosity density distribution for the disk is
(Freeman \markcite{fr70} 1970)
\begin{equation}
\rho^{(d)}(r,z)=\rho^{(d)}_0 \exp (-r/r_d) Z(z/z_d),
\label{disk}
\end{equation}

\noindent where $r$ and $z$ are the cylindrical radius and the
height above the equatorial plane of the galaxy, respectively,  and
the costants $r_d$, $z_d$ are the relative scale lengths;
The function $Z(z/z_d)$ describes the vertical behavior of
the distribution, which is assumed to be either exponential
or of the type "sech$^2$" (van der Kruit \& Searle \markcite{ks81} 1981;
Wainscoat, Freeman \& Hyland \markcite{wa89} 1989);
the results presented in this paper are for the exponential vertical
distribution that is
\begin{equation}
Z(z/z_d)=\exp (-\mid z\mid/z_d).
\end{equation}
In analogy with the bulge
case, we have introduced a horizontal, $r^{(d)}_{max}=m r_d$, and a vertical
$z^{(d)}_{max}=k z_d$ external truncation.
Our models are for $r_d = 4.0$~kpc,
$z_d= 0.35$~kpc, which are similar to what is observed for the old disk
population of the Galaxy; for the external truncations we adopt $m=k=6$.

\subsubsection{bulge/total luminosity ratios}
To simulate different Hubble types, we have run models  with different
bulge/total luminosity ratios ranging from 0 (resembling an Sd
galaxy, BT00) to 0.5 (Sa, BT05).
All our simulations are monochromatic and we have run models at the
effective wavelengths of the B and I bandpasses, $4400$ and $9000$\AA\
respectively. Possible differences in the stellar population of the two
components are taken into account by adopting a B-I color of the bulge
larger than that of the disk. More precisely the luminosity of the bulge
has been increased by one magnitude in the I band while the luminosity of
the disk has been kept constant in both bands; therefore we suppose an
intrinsic B-I=0 for the disk and B-I=1 for the bulge.

All of these prescriptions conform to those adopted by BFK.

\subsection{Dust properties}
\label{polv}

The only absorbing particles included in our code are dust grains;
we neglect the absorption above the Lyman limit due to
diffuse hydrogen gas, the possible presence of molecules, etc. .

\subsubsection{Dust disk}
The dust is supposed to be smoothly~\footnote{A study of the
effects introduced by dust clumping is in progress.} distributed in a
disk with the same law (eq.~[\ref{disk}]) as for
the stellar disk
\begin{equation}
\rho^{(g)}(r,z)=\rho^{(g)}_0 \exp (-r/r_g) Z(z/z_g).
\end{equation}
where $r_g$ and  $z_g$ are the horizontal and vertical dust scale lengths.
In this paper we are mainly interested in determining the effects of a
{\em standard} dust layer on the structural parameters and appearance
of a galaxy, we will devote a following paper to study the detailed
structure and distribution of dust as inferred from the observed light
distribution in spiral galaxies.
The Galactic parameters for the dust distribution are by
far more uncertain than for the stars. For the computations presented
here, we have assumed $r_g=4.0$~kpc, $z_g=0.14$~kpc and the same
truncation factor as for the stellar disk.
That is, while the radial extents of the stellar and dust disk are
identical, we will always deal with dust disks which are vertically
embedded into the star distribution.
This is an important point, since several of the results
in the following will be determined by the presence of sizeable
stellar emission outside the obscuring layer.

We assume that the dust physical properties (i.e. the grain
size distribution and materials, see below) are independent
of the position in the galaxy and therefore the absorption coefficient
at wavelength $\lambda$ can be written as $k_\lambda =
\langle C^{ext}(\lambda) \rangle
\rho^{(g)}(r,z) $, where $\langle C^{ext}(\lambda)\rangle$ is the  extinction
cross section
per unit mass, averaged over the grain distribution and materials (see
eq. [\ref{medie}]).

It is common to normalize the total content of dust in the galaxy
in terms of the optical depth along the simmetry axis of the galaxy,

\begin{equation}
\tau_\lambda(0)=\int_{-\infty}^{\infty}
\sigma_\lambda \rho^{(g)}(r=0,z) dz
=2\langle C^{ext}(\lambda)\rangle \rho^{(g)}_0 z_g,
\label{norma}
\end{equation}

\noindent both for the exponential and sech$^2$ distributions. As for the
stellar disk, the results here presented refer to vertically exponential
dust disks.

\subsubsection{Dust components and size distribution}
\label{compone}
In our model the grains are supposed to be spherical and to have
a size distribution given by the MRN model (Mathis, Rumpl \& Nordsiek
\markcite{ma77} 1977),
$n(a)\propto a^{-3.5}$, where $a$ is the grain radius.
We consider three materials: astronomical silicates,
$\parallel$ graphite, and $\perp$ graphite.
The numerical silicates/graphite ratio is 1:1.12,
with 1/3 of the graphite having optical properties measured parallel
($\parallel$), and 2/3 perpendicular ($\perp$) to the c-axis.
The lower and upper limits of the distribution are $a_{-}=0.005$~$\mu$m
and $a_{+}=0.25$~$\mu$m, irrespectively of the material.
Very small grains and PAH have not been included due to the large
uncertainties both in their size distribution and optical constants.

The dielectric constants adopted are the ones given by Draine \& Lee (1984),
recently extended in the far UV and X-rays by Martin \& Rouleau \markcite{ma91}
(1991). All the relevant optical properties (absorption and scattering cross
section, albedo, phase function, and Mueller matrix) have been calculated
using Mie formulae.
The resulting extinction curve is illustrated in Fig.~\ref{estinzio}.
\placefigure{estinzio}

In our models the central optical depth of the disk is defined by
\tc\, the value at 5500\AA; models at different wavelengths are still
defined by \tc\ but all computations of optical thickness are scaled
according to our extinction curve.

\subsubsection{Comparison with Henyey-Greenstein approximation}

The average value of a quantity $q$, over the dust distribution,
is here defined as

\begin{equation}
\langle q \rangle = {\sum_i w_i \int_{a_{-}}^{a_{+}}
q \;C_i^{ext}n_i(a)da\over \sum_i w_i \int_{a_{-}}^{a_{+}}
C_i^{ext}n_i(a)da},
\label{medie}
\end{equation}

\noindent where $C_i^{ext}$ is the extinction cross section, the index
$i$ referes to the material, and $w_i$ is
its weight in the distribution; both $q$ and $C_i^{ext}$ depend on
$\lambda$

Previous works dealing with scattering problems have often approximated
the phase function $\Phi$ with the Henyey-Greenstein (HG) formula
(Henyey \& Greenstein \markcite{hg41} 1941),
\begin{equation}
\Phi (\theta,\phi)= {1 \over 4\pi} {{1-g^2}\over (1+g^2-2g\cos\theta)^{3/2}},
\label{HG}
\end{equation}

\noindent where $g$ is the asymmetry parameter (that is the average value
of $\cos \theta$ weighted by the phase function); $\theta$ and $\phi$
are the polar and azimuthal scattering angles, respectively.

In Fig. \ref{duefasi} we show  a comparison
between the average phase function for
the distribution calculated with the Monte Carlo procedure implemented in
our radiative transfer code and the HG phase function for which a
value of $g$ averaged on the distribution according to
equation (\ref{medie}) has been used. The differences between the two phase
functions are striking (as an example, the B and I bands are shown):
{\it the HG approximation always underestimates the forward scattering} by
as much as 25 \%, the difference being  larger in the B band.
\placefigure{duefasi}

For the albedo and asymmetry parameter we obtain $\langle\tilde{\omega}
\rangle=0.53$ and $\langle g\rangle=0.38$ in the B-band, and $\langle
\tilde{\omega} \rangle=0.50$ and $\langle g\rangle=0.21$ in the I-band.

\section{Monte Carlo simulation}

Monte Carlo methods allow the life of each photon
to be followed
through scattering and absorption processes.  Here we provide a brief
description of our computational scheme, deferring the interested reader to the
Appendix for a more detailed presentation.

\subsection{Emission cycle}

We determine first the position of the photon emission point, according to
the bulge and disk luminosities as described in Sect. \ref{distri}.
Photons are emitted isotropically and unpolarized.
The Stokes parameters are defined for each photon on the plane
perpendicular to its propagation vector, following the conventions
in Shurcliff \markcite{sh62} (1962); the reference direction on such plane is
the projection
of the galactic $z$ axis (i.e. the simmetry axis).
All photons are emitted with the same intensity, or strength, that is with
the first Stokes parameter $\rm I=1$.
The emission cycle requires five random numbers: three for the position and
two for the direction (see Appendix~\ref{A_1}).

\subsection{Scattering cycle}
Then we determine the position of the scattering
point, i.e., the position at which the photon collides with a dust
grain; grains are assumed to be distributed as described in Sect. \ref{polv}.
This implies to determine first an optical depth $\tau$
for the scattering point,
and then the geometrical position along the line of propagation where
this $\tau$ is actually reached. It turns out that this second step is the most
consuming, in
terms of computational time, of the whole process (Appendix~\ref{A_2}).
Once the position of the scattering event is known, we pass to determine
the characteristics of the dust grain, according to the size and composition
distributions (Appendix~\ref{A_3}). Note that in principle such a
sequence renders rather straightforward
to allow for situations where the physical properties of dust vary within
the galaxy, e.g. with galactocentric distance.
These steps require two more random numbers, one to locate the diffusion point,
the other to fix the size and composition of the scatterer.
Knowing the radius and refractive index
of the grain (together with $\lambda$), it is now possible to
compute the (polar) phase function and the albedo,
and, by extracting a new random number,
to choose the (polar) angle of scattering $\theta$.
Together with the geometro-optical properties of the dust grain,
and within the frame of Mie's theory, $\theta$ determines completely
the elements of the scattering (Mueller) matrix. The matrix elements
also enter into the expression for the azimuthal phase function, so
that, by extracting a new, final random number we define completely
direction and Stokes vector of the scattered photon (see Appendix~\ref{A4}).

The updated Stokes parameters are calculated
via the following transformation
\begin{equation}
\left(
\begin{array}{c}
{\rm I'}\\
{\rm Q'}\\
{\rm U'}\\
{\rm V'}
\end{array}
\right)
\propto
{\bf R'} {\bf M} {\bf R}
\left(
\begin{array}{c}
{\rm I}\\
{\rm Q}\\
{\rm U}\\
{\rm V}
\end{array}
\right).
\end{equation}
The rotation matrix ${\bf R}$ transforms the initial Stokes parameters
into a new set whose reference direction for linear polarization lies in
the scattering plane.
${\bf M}$ is the relevant Mueller matrix with reference to the scattering
plane, which depends only on $\theta$ and on the dust properties;
the matrix  ${\bf R'}$ provides the final transformation of the Stokes
parameters to the final reference direction, that is the projection
of the $z$ axis onto the plane normal to the new direction of propagation.
In our formulation we make use of normalized Mueller matrices, the
absolute values of the Stokes parameter are adjusted according to
the incoming intensity $\rm I$ and to the albedo of the scatterer.
In summary a single scattering cycle needs the extraction of four
random numbers: one to locate the diffusion point, one to choose the
size and type of grain, and two for the direction of the scattered photon.
The scattering cycle is iterated until the photon escapes the dust layer.
or dies because its strength $\rm I$  goes below a threshold value
(${\rm I}_{ lim}=10^{-4}$).

\subsection{Output and performance}

The output of the code consists of $N_B$  maps, for each
Stokes parameter, where $N_B$  is the number of bands of
different galaxy inclination to the line of sight (Appendix~\ref{A_5}).
The models presented here are obtained with $N_B$ =15,
and each map has an extent of 201x201 pixels, with a spatial
resolution of 0.2 kpc; the resolution is therefore moderate both
in space and in inclination.

A full model (bulge+disk) with enough signal to noise ratio to distinguish
a clear polarization pattern in the outer parts of the galaxy
(with linear polarization degree of $\sim$ 1\%)
required a number of photon launches of the order
of $10^8$. Anyway, photometric accuracies comparable to those of
astronomical images are usually attainable with less than
$10^7$ launches. The code has been tested on a variety
of more simple and predictable situations: point sources within
spherical scattering envelopes, Rayleigh scattering, optically
thin regimes, absorption only, etc..

\section{ Results}
In summary, the models presented here were all obtained for: bulge
effective radius $R_e=1.6$ kpc, truncated at $15$ kpc;
disk radial scale length $r_d=4.0$ kpc, truncated at $24$ kpc;
disk exponential vertical scale length $z_d=0.35$ kpc, truncated at $2.1$ kpc;
dust radial scale length $r_g=4.0$ kpc, truncated at $24$ kpc;
dust exponential vertical scale length $z_g=0.14$ kpc, truncated at $0.84$ kpc.
All of the results obtained do not depend on the adopted truncation
factor as far as they exceed 5 scale lengths.

We ran simulations for 6 different values of \tc: 0.0, 0.5, 1.0, 2.0,
5.0, 10.0; and for 4 different B/T ratios: 0.0, 0.1, 0.3, 0.5.
The average angle~\footnote{As defined by
$\langle\theta\rangle_i=\int_{\theta_{i}^{-}}^{\theta_{i}^{+}} \theta\sin\theta
d\theta$, where $\theta_i^-$ and $\theta_i^+$ are the limiting values
for the $i$-th inclination band.} to the line of sight of the 8
independent inclination bands,
was $\langle\theta\rangle$= 20$^\circ$, 37$^\circ$, 48$^\circ$,
58$^\circ$, 66$^\circ$, 75$^\circ$, 82$^\circ$, 90$^\circ$.

All these models were produced both at 4400\AA\, for the B-band,  and at
9000\AA, for the I-band. We also ran some simulations, although not a
complete set, at 3500\AA\ (U-band).

\subsection{Extinction properties}
Since the adopted range of parameters conforms to the one used by BFK,
we can directly compare our results with theirs.
All the results about minor and major axis profiles\footnote{The shape
of the major axis profiles of systems with conspicuous bulges,
inclinations $<$80$^\circ$ and \tc$>$2, resemble closely the Type II profiles
defined by Freeman (1970). Anyway, as noted by the same author, they
often show up also in galaxies with a low dust content, such as S0's,
and therefore their explanation seems to require peculiar stellar
distributions in addition to extinction effects.}, total extinction,
face-on corrections, isophote shapes and diameters agree remarkably well
and therefore we refer to their work (for a more thorough discussion see
Bianchi \markcite{bi95} 1995). Most of differences between our
results and those of BFK are mainly imputable to the choice of the
extinction law. While BFK use the empirical data of Rieke \& Lebofsky
\markcite{ri85} (1985), that gives $A_B=1.32A_V$, we conform to the
MRN and Draine \& Lee (1984) model (Sect.~\ref{compone},
Fig.~\ref{estinzio}) and obtain $A_B=1.28A_V$.
Additional differences arise from the finite spatial resolution of our
images and from the finite resolution in inclination to the line of
sight, as explained in appendix~\ref{immaginare}.
However these differences are relevant only when intensity varies
greatly with position and inclination (the edge-on case)

We point out that the central surface brightness of disks with large
optical depth still  varies with inclination due to stars not
immersed in dust, if indeed the dust scale height is smaller than
for stars.  In addition, in the B-band, the difference between
 edge-on and  face-on central surface brightness is the same
for models with $\tc\geq2$, corresponding to 0.3 magnitudes. This
common behavior makes the study of central surface brightness
unsuitable to the determination of the optical depth of the dust disk in
spiral galaxies.
As for the influence of scattering, the main (and
obvious) result is a substantial reduction of the actual extinction for
a given \tc. As illustrated in Fig.~\ref{differe}, such reduction is
higher for greater \tc\ and lower inclination, with a maximum of
$\sim0.3$ mag for B-band models. For edge-on models, the difference is
less than 0.1 mag, so that neglecting the effect of scattering when
fitting the brightness profiles of highly inclined objects can be rather
justified (Ohta \& Kodaira \markcite{oh95} 1995).  \placefigure{differe}

\subsubsection{Local extinction}

The differences between models including scattering and models with
absorption only can also be appreciated by analyzing the local extinction.
In Fig. \ref{locE} we plot the central extinction ($r=0$) for models in
the B band ($A_B$) with $i=20^\circ$ (our inclination nearest to
face-on) as a function of the central optical depth \tc.
\placefigure{locE}
The straight line
corresponds to the Holmberg screen model, while dotted ones to
sandwich models (Disney \etal 1989) with different
$\xi$, the dust-to-stellar thickness ratio.
The left panel is for models with absorption only,
while the right panel includes the scattering: screen
and sandwiches are the same in both panels, and only include absorption.

Models with absorption only show that the behavior
of the central extinction in the pure disk models (BT00) can be
described locally by a sandwich with $\xi=0.6$, while models
including a bulge require inverse sandwiches, that is with dust
extending higher than stars from the equatorial plane ($\xi=1.1$).
This is because the bulk of the emission at the center, that is the
bulge core, is located inside the dust layer.

In models with scattering (right panel) we note that the central
extinction is reduced by several tenths of magnitude since the radiation has a
good probability to be diffused by the dust, rather than absorbed.
This is because, with our assumptions about the dust composition and
size distribution, the average albedo amounts to $\sim 0.5$, both in the
B and I-band.
Moreover the behavior of scattering models is very different from that
of sandwiches.

In Fig. \ref{locE1r0} we show the extinction at  $r=r_d$ on the major
axis, as a function of the local vertical optical depth, for the same
models as in  Figure~\ref{locE}. Since the bulge contribution is small
at this distance, all the models present nearly the same behavior,
principally regulated by the disk.
Also, models with scattering show again that sandwiches are a
poor approximation.
It is interesting to note that for $\tc<0.5$ scattering plays
such a role that extinction is basically absent.
\placefigure{locE1r0}

\subsubsection{Images and color maps}

Intensity images and B-I color maps bring additional information.
In Fig. \ref{compo} we show images in the B-band and color maps for a
bulge, a disk and a BT05 model, with inclination $82^\circ$ and \tc$=10$.
Since the extinction for bulge and disk is due to the same dust disk,
the BT05 model  is simply a superposition of the first two images.
We have used the same gray scale for the intensity images, while
color maps have different scales to evidence the details.\placefigure{compo}

Extinction is more pronunced in the bulge because stars and  dust are not
homogeneously mixed and the dust acts almost like a screen.
However, extinction also modifies the shape of the disk which
becomes asymmetric with respect to the major axis.
Color maps are positive everywhere, also in pure disks (intrinsic B-I=0),
indicating a general reddening.
The red color of the outer "halo" in the BT05
maps is due to the intrinsic (B-I=1) color of the bulge (the different
appearance is due to different gray scales).

The expected effect of scattering on colors is to render bluer the zones
of an image where a substantial fraction of emerging radiation is
contributed by forward scattering. This is due to the more asymmetric
phase function at shorter $\lambda$ (Sect.~\ref{compone}). This effect should
be more evident
in highly inclined objects and in the regions of the (dust) disk
situated between the observer and the bright bulge core (Elvius
\markcite{el56} 1956).

Analyzing the pure bulge color maps we do note a less red lane near the
center, coincident with the dust disk plane, but still with B-I$\geq 1$.
This "blue" lane cannot be accounted for by scattering effects, since it
is present also in models with absorption only.
It is due, instead, to the heavy extinction of the central parts. Where
this happens, the observed emission is mostly due to stars located out of
the dust layer, and, therefore, the reddening is small. The reddest
parts are in fact those where the local opacity (along the line of
sight) is high but less than about 15.
We note that when scattering is included, due to the lower effective
opacity for a given \tc\, this effect is reduced and the central
equatorial lane actually becames redder.
The disk presents this lane as well, but the contrast is not so strong;
as an example, we show in Fig. \ref{grad} the results along the major axis
of a pure disk with \tc=5 at various inclinations:
the right panel is for a model with scattering, while the left is for
pure absorption.
The study of the color profile has been suggested by BFK as a possible
tool to infer the (local) optical depth:  color gradients are located
in regions where the optical depth along line of sight is $\sim 1$.
Our models agree with this result.
Besides, the shape of color profiles, and then the position of
gradients, is about the same with or without scattering; this means that
pure absorption models can provide sufficiently good $\tau$ diagnostic.
\placefigure{grad}

We have also performed simulations in U-band.  Scattering effects in this
wavelength are similar to those in the B-band and, therefore, U-B color maps
are not appreciably different from those obtained with simpler absorption
models.

\subsection{Polarization analysis}

We now examine the polarization properties of
our models.  The range of parameters explored is the same as for the surface
brightness maps.

While the polarization introduced by a single
scattering is usually conspicuous, it will
be seen that the predicted degree of linear polarization is low (a few
percent).
This is due to three different causes:
\begin{enumerate}
\item Photons emerging from the same position  in a map can be either
{\em primitive}, unpolarized photons, emitted by stars and
escaped without having been intercepted by dust, or scattered, polarized
photons. The first cause of low polarization is, therefore, the dilution
by primitive radiation, which is higher in the regions of higher
surface brightness, such as the central parts of both bulges and disks.
\item If we restrict ourselves to the contribution to a certain pixel of
scattered photons only, the resultant
polarization will be maximum if the scattered photons have
similar histories. So, the
requirement for a high degree of polarization is threefold: first, the
main contribution must come from single scattering events; second,
photons must be scattered within a thin region along the line of sight;
third, the scattering region must be illuminated from a narrow solid
angle. The first and second requirements are usually met in the
parameter range of our models. The third one requires a compact stellar
distribution with respect to the dust disk and is only met by the bulge
component.
\item The polarization will obviously be highest if the scattering
grains are identical. As mentioned previously our models make use of a
distribution of grains in size and composition, and to each size and
composition our code assigns, for a given scattering angle, a
different Mueller matrix. As a result, photons emitted by the same star
and scattered at the same point in the same direction may have quite
different polarization properties,
depending on the characteristics of the scattering grain.
\end{enumerate}

The maps that we will examine are obtained by averaging the images for
the various Stokes parameters within 10x10 pixel squares. This lower
resolution has been necessary to compensate for the lower signal to
noise figures of the polarization maps.
In Fig. \ref{polexa} are collected the linear polarization maps in the B-band
at
different inclinations for the emission from the bulge only, from the
the disk only, and from a composite BT05 model. The disk of dust is the
same in all cases, namely the one with highest optical depth, $\tc =10$.
\placefigure{polexa}

The degree of polarization comes out to be higher for larger optical
depths. For the BT05 model of Fig. \ref{polexa} at 75$^\circ$ inclination, the
largest polarization degree is about 1.5\%; in the same conditions but
with $\tc = 0.5$ the maximum is 0.8\%. The zones of highest polarization
are always located on, or near, the major axis.
Another notable feature of our models
is that, while increasing \tc\, the location of the maximum polarization
moves outward; this is illustrated by the sequences in Fig. \ref{polprof}.
that  show the behavior of the polarization degree along the
major axis, with galactocentric distance for some of our models.
\placefigure{polprof}

Clearly, for the reasons discussed above, the main contribution comes
from scattering of the radiation from the bulge.
This component has a more  compact distribution with an emissivity highly
peaked at the center, which is seen by the scatterers in the disk almost
as a point source. The case of a pure bulge best illustrates the
dilution of polarization by direct radiation; the maximum degree of
polarization does not vary appreciably with inclination.

Pure disks show very little polarization when seen face-on; this fact is
mainly due to the rather flat distribution of disk stars. The
polarization degree increases with inclination, with a maximum around
50$^\circ$-80$^\circ$ and then decreases again slightly, due to the
long line of sight through the disk when approaching 90$^\circ$.
As a rule, in composite models the scattering of disk radiation will
provide the polarization of the inner regions while the bulge will
mainly polarize the external disk. Note how the strong
polarization of the outer disk in case of a pure bulge ($\approx
30\%$) is diluted by direct disk light and reduced to 1.0-1.5\%.
As a consequence, the morphology of the polarization pattern, when seen
at favorable inclinations (50$^\circ$-80$^\circ$), is different in
systems with different bulge luminosities: late spirals will have a more
uniform elliptical pattern, and early ones a two lobe structure
aligned along the major axis.

As for the direction of the linear polarization vector, it is
invariably aligned perpendicularly to the major axis. This means that, for
the scattered field, the {\bf E} component perpendicular to the plane of
scattering {\bf E$_\bot$}, is higher than the parallel one {$\bf E_\|$}
for most of the directions with a sizeable phase function. This is the
situation commonly observed in scattering by small particles, where
$x=2\pi a/\lambda \ll 1$ , as it is the case, for example, in electron
scattering. When dealing with dust grains with refractive indexes
around one, this is not the only possible situation. The pattern can be
inverted, in principle, at larger values of $x$, see for example fig. 25
of Van de Hulst \markcite{va57} (1957). We have performed a few simulations to
check if, with larger
grains or at smaller $\lambda$, the pattern would appreciably change, in
the sense of a polarization pattern with a definite alignment along the
equator, when seen edge-on. Our results indicate that, for any reasonable
dust composition and size distribution, this situation is difficult to
achieve
for $\lambda \geq 2000$~\AA. The reason is that, at such
wavelenghts, the scattered field is mainly contributed by grafite
($\bot$) grains which have a (real) refractive
index $m \geq 2$. This requires, for a radical change in the pattern
direction, values of $x \geq 1.5$, which means, in the visual
($\lambda=5500$\AA), a lower cutoff $a_{-}\geq0.13\mu$m instead of the
canonical $a_{-}=0.005\mu$m. The resulting polarization map for such a
model is illustrated in Fig.~\ref{paralle} (note the low polarization
degree).
\placefigure{paralle}

Another point to mention is the comparison between the polarization maps
and the isophotal ones: two aspects are noticeable. First, althought the
brightest, inner parts show very little polarization, measurable values
($>1\%$) are diffuse over most of the galaxy image, well inside the 25
B-mag arcsec$^{-2}$, and, therefore, accessible to observations
(Fig. \ref{osserva}).\placefigure{osserva}
Second, for a given model and inclination, we find little spatial
correlation between extinction and polarization, even in highly inclined
models with large $\tc$; for example, the equatorial dust lanes of highly
inclined disks are not strongly polarized. This is
another signature that polarization is mainly contributed by single
scattering events in the outer layers of the disk.

A clear prediction of our models concerns the wavelength dependence of
the degree of linear polarization, which we find to increase from the B
to the I-band. This appears to be a general rule, indipendently of
B/T ratios, \tc, and inclination.
Such effect is ascribed to the reduced extinction in the I-band and
the subsequent increase of the number of single scatterings.
Most of the effects mentioned above are clearly depicted in
Fig.~\ref{grad}.
The positions of the maxima coincide with regions  where the local optical
depth is $\sim 1$, similarly to what previously observed for color gradients.

If the number of scattering is larger than one, Mie scattering can in
principle introduce a certain degree of local circular polarization. In
all our model the circular polarization degree is contained within the
noise level and never shows at levels higher then 0.1\%.

\section{Summary and conclusions}

To study the extinction and polarization properties of  dusty spiral galaxies,
we have performed extensive Monte Carlo simulations using realistic galaxy
models that include both a bulge and a disk component, simulating different
Hubble types. The scattering properties are calculated from Mie's theory
for each dust grain; in addition, we follow the radiative transfer of the
four Stokes parameters. The grains are assumed to
be spherical, with a power-law size distribution, and consisting either of
graphite
or silicate. This enables us to produce
both photometric and polarimetric maps of  the galaxy for different optical
depths, inclinations and B/T ratios; the main results derived
from the analysis of such maps are summarized in the following.

In general, we find a very good agreement with the results of BFK, who
used a series of approximations described above, concerning luminosity
profiles, total extinction, isophote shapes and diameters.

We have shown that the overall effect of scattering (with respect to
pure absorption) is to reduce the
extinction for a given \tc. Such decrement in the effective $A_B$ is
particularly evident in the central parts of bright bulges, as shown in
Fig.~\ref{locE}, although the extinction fractional reduction is larger
for late-type than for early-type spirals and increases with
galactocentric distance. Sandwich models, often used in the literature,
are shown to fail dramatically in reproducing the extinction - optical depth
relation when scattering is properly considered, and they should not be used.
{}From the analysis of the B-I color maps we conclude  that there is a net
reddening of the galaxy due to dust even when scattering is included.
In fact, the latter reduces on average the B-I color by about 0.1~mag, a
mild but appreciable effect.
Radial color gradients are found to exist in regions where the
optical depth is of order unity.

The degree of linear polarization produced by scattering is usually small
(of the order of a few percent), but still practically measurable.
The reasons for this behavior are: (i) there is a dilution effect by
direct stellar (unpolarized) radiation; (ii) scattering region is
usually illuminated from a large solid angle, thus photons with different
histories enter the same beam, and (iii) the distribution of grain
sizes and materials degrades the coherence of the polarization.
Most of the polarization comes from light from the bulge which is
single scattered by the outer layers of the disk. This is not
surprising, of course, since the bulge is more compact and resembles a
point source. The polarization degree increases with the optical depth
and with inclination, flattening out above $\sim 80^o$. The
polarization pattern is found to be perpendicular to the major axis,
independently of the morphology,  inclination and optical depth; this
result depends only on the optical properties of the dust grains, and
can be reversed only by shifting to much larger radii the lower limit
for the dust size.

It is interesting to note that some galaxies do show polarization patterns
similar to those of our models
(Scarrott, \markcite{sc90} Rolph, \& Semple 1990;
Fendt et al. \markcite{fe95} 1995). Among
them, late type galaxies (NGC 5907, NGC 891) show polarization
perpendicular to the major axis on the entire extent of the disk;
early type galaxies (NGC 4594, NGC 4565), instead, show such
polarization in the outer
regions of the disk, whereas the central parts have polarization
vectors consistent with dichroic absorption, that is parallel to the
major axis.
When polarization vectors are perpendicular to the major axis,
the observed degree of linear polarization is about one percent,
in agreement with our models.

Finally, there is little correlation between
extinction and polarization degree and there is a trend of increasing
polarization from the B to the I band due to the reduced extinction in
the I band and subsequent increase of  single scatterings.

\acknowledgments
Part of this study has been supported by the Programma Vigoni.
We acknowledge
several stimulating discussions with R.-J. Dettmar, F. Menard, E. Landi
degl'Innocenti and M. Landolfi.
One of us (S. B.) wishes to thank the Arcetri Chaos team for useful
suggestions.

\appendix
\section{Description of the Monte Carlo code}

With the Monte Carlo method each event in the photon life is determined by
the probability distribution of a pseudo
random variable, $\rho(t)$, that describes
the statistical weight of the variable $t$ in the event, and by
extracting a random number $R\in[0,1]$. The value $t^\star$
associated to the random number $R$ according to
$\rho(t)$ is given by the general Monte Carlo formula
\begin{equation}
\frac{\displaystyle \int_{t_{min}}^{t^\star}\rho(t)dt}
{\displaystyle \int_{t_{min}}^{t_{max}}\rho(t)dt}=R
\label{A1}
\end{equation}
where $t_{min}$ and $t_{max}$ define the interval of possible values for
$t$.  In the following $R_i$ will indicate a random number between $0$ and
$1$.

As a standard reference frame we adopt a cartesian one, having the $z$
axis along the galactic symmetry axis, the $x$ and $y$ axes on the
equatorial plane. The $y$ axis coincides with the nodal line
(intersection between sky plane and equatorial plane).

\subsection{Emission}
\label{A_1}
In our models photons are emitted by stars distributed in the galactic
disk and bulge. The first step is to determine the  coordinates of
the star (i.e. the photon emission point) according to the star density
distributions.

\subsubsection{Bulge emission}
\label{asas}
Using the Jaffe law, and
substituting equation (\ref{jaffe}) in equation (\ref{A1}), one
finds a simple relation between the radial distance and a random number
$R_1$
\begin{equation}
\tilde{r}^\star=r_b\frac{R_1}{\frac{n+1}{n}-R_1}
\label{A2}
\end{equation}
where $n$ is the truncating factor of the Jaffe bulge and $r_b$ the
scale radius.
The $R^{1/4}$ and Jaffe distributions are
relatively similar, but the relation between $r_b$ and $R_e$ depends on which
physical parameters, and parts of the galaxy, we intend to match best. To our
purpose, we need the relation fitting the luminosity
within a given $\tilde{r}$, since this is the physical variable which we will
directly associate with the Monte Carlo random numbers (this will be the case
for the disk radial distribution as well). In this sense, when uniformly
sampled
over the luminosity content, the best correspondence between the two
distributions is found for $r_b = 1.16\;R_e$.

Then we determine the polar, $\theta$, and azimuthal, $\phi$,
angles in spherical coordinates.  With spherical symmetry we have:
\begin{equation}
\begin{array}{ccc}
\starred{\phi}=2\pi R_2,&\;\;\;\;&
\starred{\theta}=\cos^{-1}(1-2R_3),
\end{array}
\label{A3}
\end{equation}
where $R_2$ and $R_3$ are two indipendent random numbers.
Thus, the emission coordinates are
\begin{equation}
\begin{array}{ccc}
x_0&=&\tilde{r}^\star\sin\starred{\theta}\cos\starred{\phi}\\
y_0&=&\tilde{r}^\star\sin\starred{\theta}\sin\starred{\phi}\\
z_0&=&\tilde{r}^\star\cos\starred{\theta}.
\end{array}
\label{A4}
\end{equation}
In the more general case of axially symmetric ellipsoids, $z_0$ becomes
\begin{equation}
z_0=\tilde{r}^\star\epsilon\cos\starred{\theta}
\end{equation}
where $\epsilon$ is the ratio between the minor and major axis.

\subsubsection{Disk emission}
For an exponential disk (eq. [\ref{disk}]), the cylindrical radius of emission
$\starred{r}$ can be found inverting numerically
\begin{equation}
\frac{\displaystyle \int_{0}^{\starred{r}}e^{-\frac{r}{r_d}}
rdr}{\displaystyle \int_{0}^{r_{max}}e^{-\frac{r}{r_d}}rdr}=
\frac{1-(1+\starred{t})e^{-\starred{t}}}
{1-(1+m)e^{-m}}=R_1'
\label{A5}
\end{equation}
with $\starred{t}=\starred{r}/r_d$.
{}From the azimuthal simmetry around the {\em z} axis:
\begin{equation}
\starred{\phi}=2\pi R_2'.
\end{equation}
We calculate  $z$ using
\begin{equation}
\frac{\displaystyle \int_{-z_{max}}^{\starred{z}}Z(\frac{z}{z_d})dz}
{\displaystyle \int_{-z_{max}}^{z_{max}}Z(\frac{z}{z_d})dz}=R_3'
\end{equation}
where
\begin{equation}
Z(\frac{z}{z_d})=
\left\{
\begin{array}{c}
\exp (-\mid z \mid / z_d),\\
\mbox{sech}^2\left(z/z_d\right).
\end{array}
\right.
\end{equation}
For the exponential law, we have
\begin{equation}
\starred{t}=-\mbox{sgn}(R_3' -\frac{1}{2})\ln\left[\mbox{sgn}(R_3'
-\frac{1}{2})(1-2R_3')(1-e^{-k})+1\right];
\end{equation}
for the $\mbox{sech}^2$ law,
the relation is
\begin{equation}
\starred{t}=\tanh^{-1}\left\{2\left[\tanh(k)\right]
R_3' - \tanh(k)\right\},
\end{equation}
where $\starred{t}=\starred{z} / z_d$ and $k$ is the vertical truncation
factor.  Thus
\begin{equation}
\begin{array}{ccc}
x_0&=&\starred{r}\cos\starred{\phi}\\
y_0&=&\starred{r}\sin\starred{\phi}\\
z_0&=&\starred{z}.
\end{array}
\end{equation}

\subsubsection{Direction}
If the radiation is emitted isotropically by stars, the longitude and
colatitude of its direction are
\begin{equation}
\begin{array}{ccc}
\phi=2\pi R_4,&\;\;\;\;\;&
\theta=\cos^{-1}(1-2R_5),
\end{array}
\end{equation}
from which the direction cosines are computed
\begin{equation}
\begin{array}{ccc}
l_0&=&\sin\theta\cos\phi\\
m_0&=&\sin\theta\sin\phi\\
n_0&=&\cos\theta.
\end{array}
\end{equation}
We define the unit vector  ${\bf v}_0 \equiv (l_0,m_0,n_0)$.

The stellar radiation is largely unpolarized, so we will assume for
the initial Stokes parameters
\begin{equation}
\left(
\begin{array}{c}
{\rm I_0}\\
{\rm Q_0}\\
{\rm U_0}\\
{\rm V_0}
\end{array}
\right)=
\left(
\begin{array}{c}
1\\
0\\
0\\
0\end{array}
\right).
\label{A6}
\end{equation}
In the following the Stokes parameters will be calculated assuming as
reference direction the projection of the $z$ axis on the plane
normal to the photon direction (for unpolarized radiation equation [\ref{A6}]
is true in all reference frames).

In summary, for the emission of every photon, we need to extract five random
numbers, three  for the coordinates of the point of emission and two for the
direction of emission.

\subsection{Diffusion point}
\label{A_2}
A photon emitted in ($x_0$,$y_0$,$z_0$) along the direction ${\bf v}_0$
travels undisturbed until it collides with a dust grain.
If undisturbed, the geometrical path within the dust distribution would
be:
\begin{equation}
\begin{array}{cccc}
\Gamma(t)&=&\left\{\begin{array}{ccl}
      x&=&x_0+l_0 t\\
      y&=&y_0+m_0 t\\
      z&=&z_0+n_0 t
      \end{array}\right.&0\leq t <t_T
\end{array}
\label{A7}
\end{equation}
where $t_T$ is the t-value corresponding to the point of exit from
the dust disk. The total optical depth $\tau_T$ along this path is
\begin{equation}
\tau_T=\frac{\tau_\lambda(0)}{2 z_g}\int_0^{t_T}
\frac{\rho^{(g)}(r,z)}{\rho_0^{(g)}} dt
\label{A8}
\end{equation}
where we made use of equation (\ref{norma}), and $r=\sqrt{x^2+y^2}$.
A fraction $e^{-\tau_T}$ of the photons will then emerge undisturbed from
the system, while the remaining $1-e^{-\tau_T}$ will be absorbed or
diffused in other directions.
The point of diffusion is associated with a random number $R_6$, which
determines the total optical depth at the diffusion point according to
\begin{equation}
\int_0^{\tau}e^{-\xi}d\xi=R_6,
\label{e1}
\end{equation}
or
\begin{equation}
\tau=-\ln(1-R_6).
\label{e2}
\end{equation}
If $\tau$ is greater than $\tau_T$, the photon will exit without
scattering.  This procedure, however, is quite inefficient: if $\tau_T$ is
small, most of the photons will exit directly. To overcome this problem, we
have used the method of {\em forced scattering} introduced by Cashwell
\& Everett \markcite{ca59} (1959; see also Witt \markcite{wi77}
1977). With this method the photon is split into two components: one,
with intensity $I_1=\exp ( -\tau_T)$, keeps travelling undisturbed
and exits the dust disk, the other, with intensity $I_2=1-\exp (
-\tau_T)$ is forced to scatter. Accordingly, equations (\ref{e1}) and
(\ref{e2}) become
\begin{equation}
\frac{\displaystyle \int_0^{\tau}e^{-\xi}d\xi}
{\displaystyle \int_0^{\tau_T}e^{-\xi}d\xi}=R_6.
\end{equation}
and
\begin{equation}
\tau=-\ln(1-R_6(1-e^{-\tau_T})),
\end{equation}
respectively.
In the  limit $\tau_T\rightarrow\infty$, $\tau$ tends to the value for
unforced scattering. The  other Stokes parameters are scaled
accordingly. Forcing the scattering has also the effect of reducing
the noise in the resulting images, for the same total number of photons
launches, but it is appreciably more time consuming. We performed a
series of trial simulations using a point source embedded in spherical
envelopes of optical opacity from 1 to 10. As a result of these tests,
we adopted the following scheme in our code:
the first scattering is always forced;
the following ones are forced only if $\tau_T>1$. Among the various
tested schemes, this one attains the best S/N ratio within a given
computational time.

We have used a threshold value $\tau_{lim}$ for the optical depth:
if a photon travels along a path with $\tau_T<\tau_{lim}$ it is allowed
to escape without being attenuated; the adopted value is  $\tau_{lim}=10^{-4}$.
Once $\tau$, the optical depth of the impact event, has been calculated, the
integral in equation (\ref{A8}), but with $t_1$ instead of $t_T$,
is inverted  to find the value of $t_1$, the geometrical length
travelled before impact.
The coordinates of the first scattering are then
\begin{equation}
\begin{array}{ccc}
x_1&=&x_0+l_0 t_1\\
y_1&=&y_0+m_0 t_1\\
z_1&=&z_0+n_0 t_1.
\end{array}
\end{equation}

\subsection{Choice of the dust grain}
\label{A_3}

We now need to know the geometrical and optical characteristics of the
dust particle located at $x_1$,$y_1$,$z_1$.
The probability for a photon to collide with a spherical grain of
radius $a$, refractive index $m$, and material $i$, is proportional to
the product of the extinction cross section, ${\rm C}_i^{ext}$,
and the grain number density $n(a)\propto a^{-3.5}$ (MRN model).
The differential collision probability  can be written as
\begin{equation}
dP_i(a)\propto w_i a^{-3.5} {\rm C}^{ext}_i da,
\end{equation}
where  $w_i$ is the statistical weight of material $i$.
Let us call $\chi_i(a)$ the function
\begin{equation}
\displaystyle
\begin{array}{cc}
\chi_i(a)= {\displaystyle \int_{a_{-}}^a} dP_i(a)&\mbox{$a\in[a_{-},a_{+}]$.};
\end{array}
\end{equation}
$\chi_i(a)$ can be normalized to unity dividing it by the value
$\chi_{tot}(a_{+})=\sum_i \chi_i(a_{+}) $:
\begin{equation}
\overline{\chi}_i(a)=\frac{\chi_i(a)}{\chi_{tot}(a_{+})}.
\end{equation}
$\overline{\chi}_i(a)$ has now values between 0 and
$\overline{\chi}_i(a_{+})$.
The interval $[0,1]$ can be divided in three subintervals for the three
materials (Si, C$\parallel$, C$\perp$)
\begin{equation}
\begin{array}{c}
\left[0,\overline{\chi}_{\rm Si}(a_{+})\right],\\\\
\left[\overline{\chi}_{\rm Si}(a_{+}),
\overline{\chi}_{\rm Si}(a_{+}) + \overline{\chi}_{\rm C\parallel} (a_{+})
\right],\\\\
\left[ \overline{\chi}_{\rm Si}(a_{+}) +
\overline{\chi}_{\rm C\parallel}(a_{+}),1\right].
\end{array}
\end{equation}
In this way, by means of a random number $R_7$ we choose both the
grain material and size. The grain size is found inverting the integral
\begin{equation}
\int_{a_{-}}^{a} \overline{\chi}_i(a)da + B_i=R_7,
\label{A9}
\end{equation}
where $B_i$ is the value  associated with the minimum value of each
subinterval ($B_{\rm Si}=0$, $B_{\rm C\parallel}=\overline{\chi}_{\rm
Si}(a_{+})$, etc.).
By knowing the wavelength and the material, we determine the complex
refractive index, which, together with size, characterizes completely
the optical properties of the grain.

\subsection{Scattering direction and polarization transfer}
\label{A_4}
If {\bf v}$_0$ and {\bf v}$_1$ are unit vectors defining the propagation
direction before and after scattering, respectively, we define
{\bf p}$_{z0}$ as the unit vector
indicating the projection of the positive $z$ axis (the galactic
symmetry axis specified by the unit vector {\bf k}) onto the plane
perpendicular to {\bf v}$_0$ at the
scattering point.
As mentioned before, {\bf p}$_{z0}$ defines the reference direction for
the Stokes parameters of the incoming photon.

Two angles define {\bf v}$_1$ with respect to {\bf v}$_0$
(Fig. \ref{v0_pz0}): the polar angle, $\theta$, between {\bf v}$_0$ e
{\bf v}$_1$, which is the actual angle of scattering, and the azimuthal
angle, $\phi$, between {\bf p}$_{z0}$ and {\bf s}$_0$
the projection of {\bf v}$_1$ on the plane normal to {\bf v}$_0$.
The vector {\bf s}$_0$ also lies in  the scattering plane
defined by {\bf v}$_{0}$ and {\bf v}$_{1}$.
\placefigure{v0_pz0}

\noindent After some algebra we derive
\begin{equation}
{\bf p}_{z0}=\frac{1}{\sqrt{1-n_0^2}}\left(-l_0n_0,-m_0n_0,1-n_0^2\right)
\label{A10}
\end{equation}

Since the Mueller matrix elements are more easily defined for Stokes parameters
referred to the scattering plane, we first have to perform a rotation
about {\bf v}$_{0}$ and change to {\bf s}$_0$ the reference direction.
The Stokes parameters of the incoming photon referred to the scattering
plane are then
\begin{equation}
\begin{array}{cccc}
\left(\begin{array}{c}
{\rm I}_0'\\{\rm Q}_0'\\{\rm U}_0'\\{\rm V}_0'
\end{array}\right)
&=&\left(
\begin{array}{cccc}
1&0&0&0\\
0&\cos2\phi&\sin2\phi&0\\
0&-\sin2\phi&\cos2\phi&0\\
0&0&0&1
\end{array}\right)&\left(\begin{array}{c}
{\rm I}_0\\ {\rm Q}_0\\ {\rm U}_0\\ {\rm V}_0\end{array}\right)
\end{array}
\end{equation}
With our assumptions, the phase function for polarized radiation is (van
de Hulst 1957, pg. 15)
\begin{eqnarray}
\Phi\left(\theta,\phi\right)&=&\frac{1}{\pi x^2 {\rm Q}^{sca}}\left[
{\rm S}_{11}+{\rm S}_{12} \frac{{\rm Q}_0'}{{\rm I}_0'} \right]
\nonumber \\
\nonumber \\
&=&\frac{1}{\pi x^2 {\rm Q}^{sca}}
\left[{\rm S}_{11}\left(\theta\right)+
      {\rm S}_{12}\left(\theta\right)
      \left(
      \frac{{\rm Q}_0}{{\rm I}_0}\cos2\phi +
      \frac{{\rm U}_0}{{\rm I}_0}\sin2\phi
      \right)
      \right].
\label{A11}
\end{eqnarray}
where
\begin{equation}
\begin{array}{ccc}
x=\frac{2\pi a}{\lambda},&\;\;\;\;&
Q^{sca}=\frac{C^{sca}}{\pi a^2}.
\end{array}
\end{equation}
${\rm S}_{11}$ and  ${\rm S}_{12}$ are two elements of the
Mueller matrix, which are functions only of $\theta$ (spherical grains),
and ${\rm C}^{sca}$ is the scattering cross section (analogous to ${\rm
C}^{ext}$).
The angles $\theta$ e $\phi$ are distributed according to equation (\ref{A11}).
The phase function, $\Phi$, is normalized, so that
\begin{equation}
1 = \int_0^{2\pi} d\phi \int_0^{\pi}d\theta\sin\theta
\;\Phi\left(\theta,\phi\right) =
 \int_0^{\pi}d\theta\sin\theta \;\tilde{\Phi}\left(\theta\right)
\end{equation}
where $\tilde{\Phi}$ is
\begin{equation}
\tilde{\Phi}(\theta)=\int_0^{2\pi}\Phi\left(\theta,\phi\right)d\phi=
\frac{2}{x^2 {\rm Q}^{sca}}{\rm S}_{11} \left(\theta\right);
\end{equation}
$\tilde{\Phi}$ depends only on $\theta$ so that $\theta$ can be derived
from the Monte Carlo method, indipendently both of
$\phi$ and of the polarization of the incident radiation.
The scattering polar angle, $\overline{\theta}$, can be
calculated from a random number $R_8$, inverting
\begin{equation}
\int_0^{\overline{\theta}} \tilde{\Phi}(\theta)\sin\theta d\theta=R_8.
\label{A12}
\end{equation}
Substituting this value of  $\overline{\theta}$ in equation (\ref{A11})
we can find $\overline{\phi}$, from
\begin{equation}
\frac{\displaystyle \int_0^{\overline{\phi}}\Phi(\overline{\theta}
,\phi)d\phi}
{\displaystyle \int_0^{2\pi}\Phi(\overline{\theta},\phi)d\phi}=R_9,
\end{equation}
or
\begin{equation}
\displaystyle \frac{1}{2\pi}\left[
\overline{\phi} +\frac{{\rm S}_{12}\left(\overline{\theta}\right)}
{2{\rm S}_{11}\left(\overline{\theta}\right)}\left(\frac{{\rm Q}_0}{{\rm
I}_0}\sin2\overline{\phi}+\frac{{\rm U}_0}{{\rm I}_0}\left(1-\cos2
\overline{\phi}\right) \right)\right]=R_9.
\label{A13}
\end{equation}

\noindent Using the values of $\overline{\theta}$ and $\overline{\phi}$
obtained before, we compute the direction cosines of {\bf v}$_1$
\begin{eqnarray}
l_1&=&\frac{\sin\overline{\theta}}{\sqrt{1-n_0^2}}
\left(-l_0 n_0\cos\overline{\phi}+m_0\sin\overline{\phi}\right) +
l_0\cos\overline{\theta} \nonumber \\
m_1&=&\frac{-\sin\overline{\theta}}{\sqrt{1-n_0^2}}
\left(m_0 n_0\cos\overline{\phi}+l_0\sin\overline{\phi}\right) +
m_0\cos\overline{\theta} \nonumber \\
n_1&=&\sqrt{1-n_0^2}\sin\overline{\theta}\cos\overline{\phi} +
n_0\cos\overline{\theta}.
\end{eqnarray}

\noindent The new direction of propagation allow us to calculate the updated
Stokes parameters after scattering (but still referred to the scattering
plane) by applying the Mueller matrix
\begin{equation}
\begin{array}{cccc}
\left(
\begin{array}{c}
{\rm I_1''}\\{\rm Q_1''}\\{\rm U_1''}\\{\rm V_1''}
\end{array}
\right)
&
=
&
\left(
\begin{array}{cccc}
{\rm S}_{11}&{\rm S}_{12}&0&0\\
{\rm S}_{12}&{\rm S}_{11}&0&0\\
0&0&{\rm S}_{33}&{\rm S}_{34}\\
0&0&-{\rm S}_{34}&{\rm S}_{33}
\end{array}
\right)
&
\left(
\begin{array}{c}
{\rm I_0'}\\{\rm Q_0'}\\{\rm U_0'}\\{\rm V_0'}
\end{array}
\right)
\end{array}.
\label{mutra}
\end{equation}
It will be noted that equation (\ref{mutra}) is different from the usual
relation for the scattered Stokes vector, since it does not include the
spherical factor $1/k^2 r^2$. The standard formulae (e.g. Bohren \&
Huffman \markcite{bo83} 1983, eq. [3.16]) are derived for
the distribution of the scattered electromagnetic fields, within the
frame of the electromagnetic theory. In our
description, the scattered photon is not distributed over the entire
solid angle but, instead, deflected onto a new single direction, and
attenuated by a factor equal to the albedo (which in Mie's approximation
is independent of direction). If I$_1'$ is the correct value of the
scattered intensity, we have:
\begin{equation}
{\rm I_1'}=\tilde{\omega} {\rm I_0'}=\tilde{\omega} {\rm
I_0},
\end{equation}
and the correct value of the Stokes vector after scattering will be
\begin{equation}
\left(
\begin{array}{c}
{\rm I_1'}\\{\rm Q_1'}\\{\rm U_1'}\\{\rm V_1'}
\end{array}
\right) = \frac{\tilde{\omega} {\rm I_0}}{\rm I_1''}\left(
\begin{array}{c}
{\rm I_1''}\\{\rm Q_1''}\\{\rm U_1''}\\{\rm V_1''}
\end{array}\right).
\end{equation}
The new  Stokes parameters are defined with respect to {\bf s}$_1$,
the unit vector which lies in the scattering plane and in the plane
perpendicular to the new direction of propagation {\bf v}$_1$
(Fig. \ref{polare}): coherently with the convention adopted, we now
have to calculate the Stokes parameters using, as a reference, {\bf p}$_{z1}$
the projection of the $z$ axis onto the plane normal to {\bf v}$_1$.
\placefigure{polare}

The expression for {\bf p}$_{z1}$ is analogous to that for {\bf
p}$_{z0}$, provided we substitute {\bf v}$_1$, for {\bf v}$_0$. Since
{\bf s}$_1$ lies in  the scattering plane,
it is parallel to the projection of {\bf v}$_0$ onto the plane normal
to {\bf v}$_1$ (but with opposite direction), and we have
\begin{equation}
{\bf s}_1\times\left(
{\bf v}_0-(
{\bf v}_0\cdot{\bf v}_1
){\bf v}_1\right)=0
\end{equation}
{}From the previous relation we have
\begin{equation}
{\bf s}_1=\frac{1}{\sqrt{1-({\bf v}_0\cdot{\bf v}_1)^2}} \left(
l_1({\bf v}_0\cdot{\bf v}_1)-l_0,
m_1({\bf v}_0\cdot{\bf v}_1)-m_0,
n_1({\bf v}_0\cdot{\bf v}_1)-n_0\right)
\end{equation}
The angle $\gamma$ between {\bf p}$_{z1}$ and {\bf s}$_1$ is found using
two other relations:
\begin{equation}
\cos\gamma={\bf p}_{z1}\cdot{\bf s}_1=\frac{n_1({\bf v}_0\cdot{\bf
v}_1)-n_0}{\sqrt{1-n_1^2}\sqrt{1-({\bf v}_0\cdot{\bf v}_1)^2}}
\end{equation}
\begin{equation}
\sin\gamma{\bf v}_1=
{\bf p}_{z1}\times{\bf s}_1=
\frac{m_0 l_1-l_0 m_1}
{\sqrt{1-n_1^2}\sqrt{1-({\bf v}_0\cdot{\bf v}_1)^2}}{\bf v}_1.
\end{equation}
The Stokes parameters  referred to {\bf p}$_{z1}$ will be calculated
applying a rotation by an angle (-$\gamma$)
\begin{equation}
\begin{array}{cccc}
\left(\begin{array}{c}
{\rm I}_1\\{\rm Q}_1\\{\rm U}_1\\{\rm V}_1
\end{array}\right)
&=&\left(
\begin{array}{cccc}
1&0&0&0\\
0&\cos2\gamma&-\sin2\gamma&0\\
0&\sin2\gamma&\cos2\gamma&0\\
0&0&0&1\end{array}\right)&\left(\begin{array}{c}
{\rm I}_1'\\ {\rm Q}_1'\\ {\rm U}_1'\\ {\rm V}_1'\end{array}\right)
\end{array}
\end{equation}

In summary, every diffusion event requires four random numbers, one
to locate the diffusion point, one for the optical characteristics of the
grain and two to determine the new direction of propagation.

\subsection{Imaging}
\label{A_5}
\label{immaginare}

After $N$ scatterings, the photon  parameters satisfy the
exit conditions.
The final parameters are the point of last scattering, $x_N$,$y_N$,$z_N$,
 the exit direction, {\bf v}$_N$=($l_N$,$m_N$,$n_N$), and the
exit Stokes parameters, I$_N$,Q$_N$,U$_N$,V$_N$.

Our galactic models have two simmetries that can be used to reduce
the computational time: (i) axial simmetry around the $z$ axis and (ii)
planar simmetry about the equatorial ($xy$) plane.
If we look at the galaxy from a point at infinite distance in
the  $xz$ plane, given the axial simmetry, we
perform a rotation around the $z$ axis and align {\bf v}$_n$ parallel to
the $xz$ plane.
After rotation, the exit direction can be identified by the new
coordinates of the point of last scattering ($x$,$y$,$z$)
and by the polar angle $\theta$ only (we omit pedix N for simplicity).
The corresponding Stokes parameters (I,Q,U,V) will be the same as
I$_n$,Q$_n$,U$_n$,V$_n$ because a rotation around the $z$ axis does not
modify the relative orientation of the wave electric field and the $z$ axis.
Because of the axial simmetry, for each photon exiting from
($x$,$y$,$z$) with direction $\theta$  and parameters
(I,Q,U,V) we can add another photon exiting from ($x$,$-y$,$z$)
with the same direction and parameters (I,Q,-U,-V).
We can then exploit the planar simmetry: for each photon exiting from
($x$,$y$,$z$) with direction given by $\theta$  and parameters (I,Q,U,V)
we add a photon with ($x$,$y$,$-z$), direction $\pi-\theta$
and parameters (I,Q,-U,-V).
Using these simmetries we gain, therefore, a factor four in the number
of photons.

To produce maps of the galaxy as seen from different inclinations we
select the photons, after their exit, according to their $\theta$
direction. This is done by dividing the whole solid angle in $N_{\rm b}$
latitudinal bands, of equal solid angle. All the photons with exit
direction within a given band will contribute to the image relative to that
range of $\theta$ values.
Note that this introduces a finite resolution in inclination: for
example with our standard ($N_B$=15) sampling, the most face-on image
includes inclinations ranging from 0$^\circ$ to 30$^\circ$, while the
edge-on case includes inclinations between 86$^\circ$ and 94$^\circ$.
Having already exploited the planar symmetry, images with inclination
$\theta$ and $\pi-\theta$ will be identical.
Finally, the photons pertaining to a certain inclination band are
projected onto the plane of the sky, according to their point of last
scattering and direction.

The final result consists of $N_B$ images for each Stokes parameter.
Linear polarization maps are obtained calculating, for each pixel, the
linear polarization,
\begin{equation}
P=\frac{\sqrt{{\rm Q}^2+{\rm U}^2}}{\rm I},
\end{equation}
and the polarization angle,
\begin{equation}
\psi=\frac{1}{2}\mbox{atan}\frac{\rm U}{\rm Q}+\psi_0
\end{equation}
where $\psi_0$ is defined in Calamai, Landi degl'Innocenti, \&Landi
degl'Innocenti\markcite{ca75} (1975).

\newpage
\figcaption[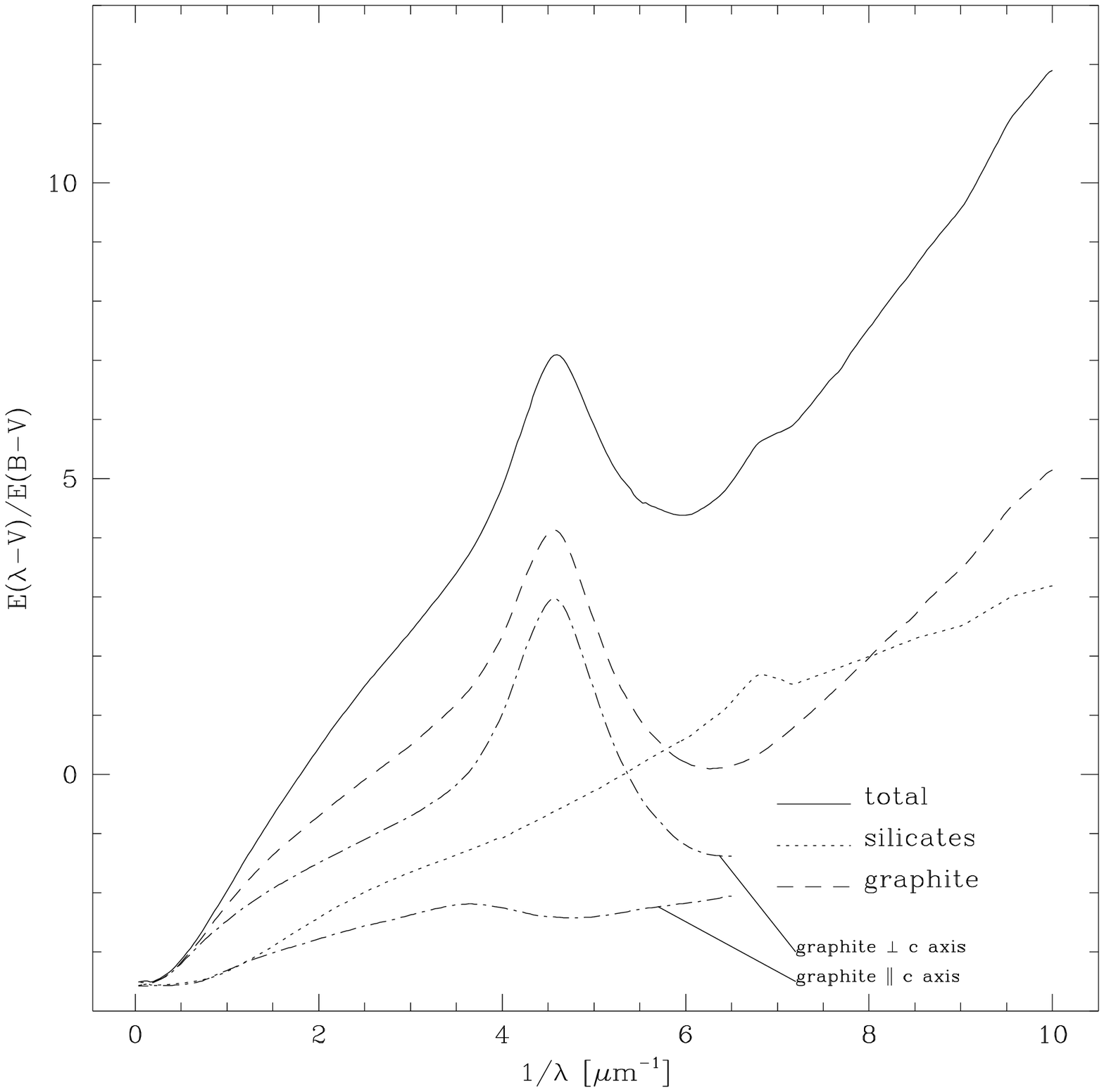]{The extinction curve resulting from the adopted
dust model. Also shown are the contributions of the single components.
The color excesses for the single materials are divided by the value of
E(B-V) for the total mixture. \label{estinzio}}

\figcaption[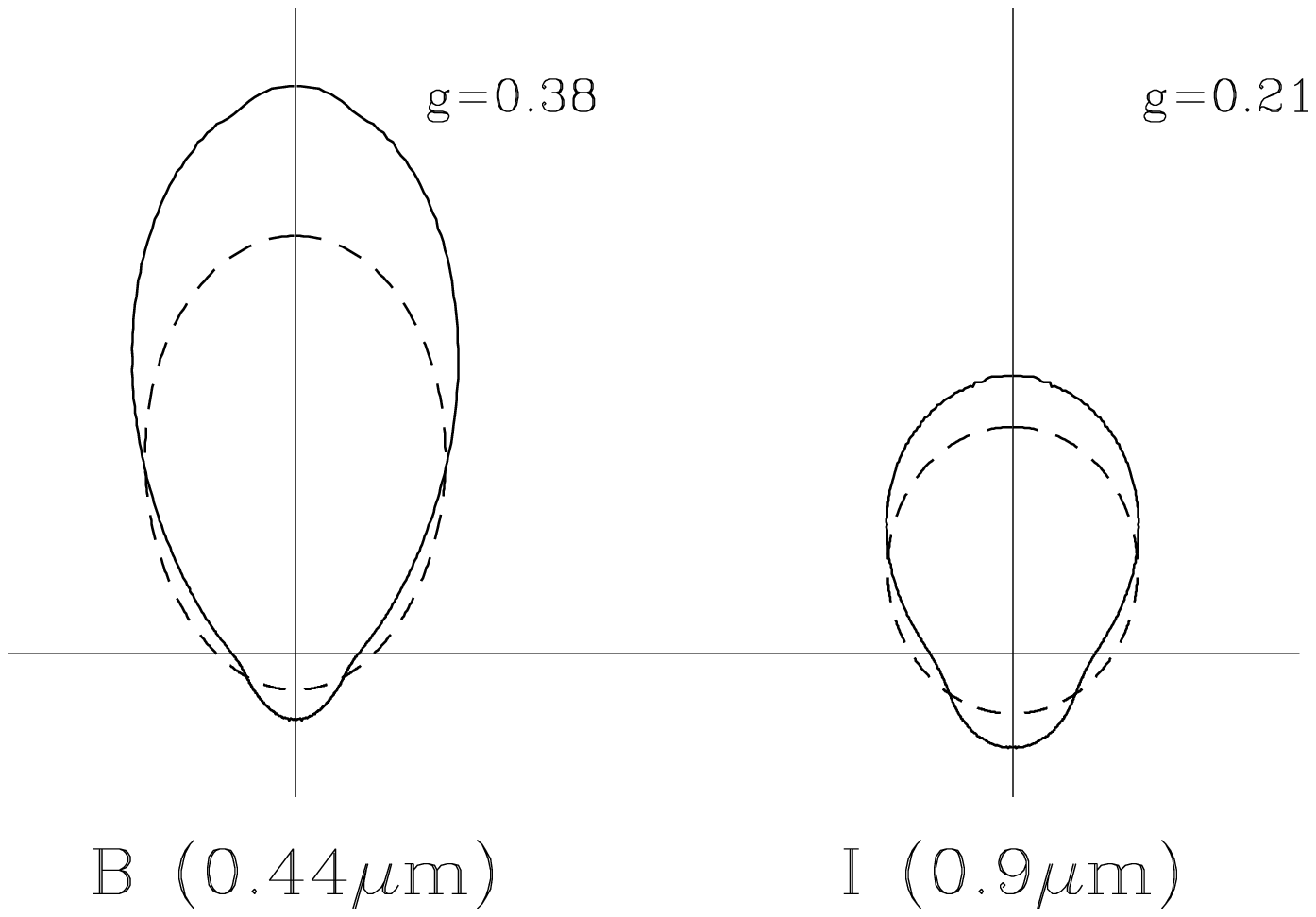]{ Average phase functions as computed by the Monte
Carlo code (solid  lines) and HG phase functions with $g$
corresponding to the mean value for the same dust distribution
(dashed lines); the phase functions are shown at the central wavelength
of the B and I bandpasses. \label{duefasi}}

\figcaption[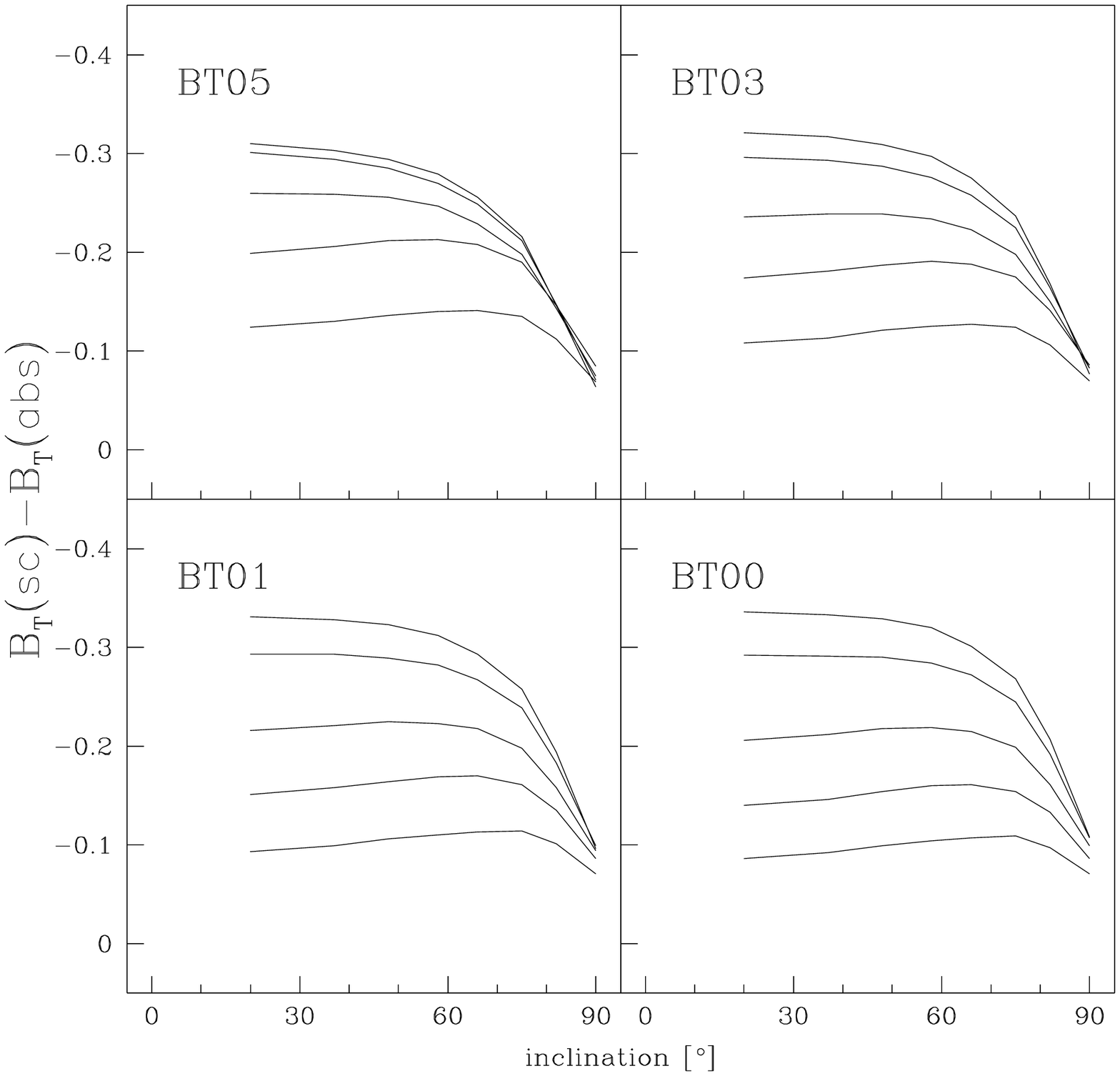]{Difference in B total magnitude between models
including scattering, $B_T({\rm sc})$, and with absorption only,
$B_T({\rm abs})$,
as a function of inclination, for different bulge/total ratios. In each
panel the five lines refer to different total opacities: $\tc=$ 10, 5,
2, 1, 0.5, from top to bottom.\label{differe}}

\figcaption[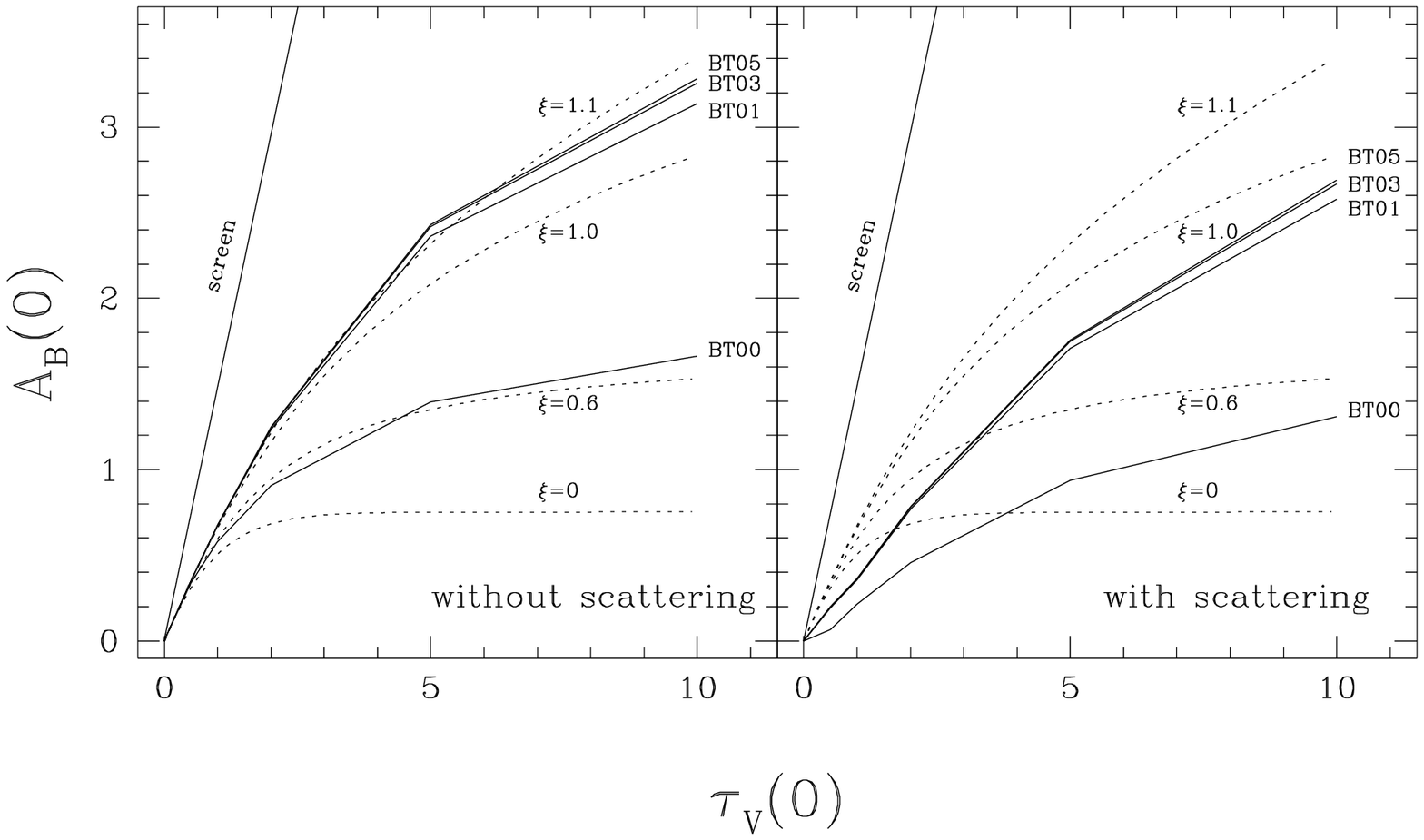]{ Central ($r=0$) extinction versus optical depth \tc\
for B models with $i=20^\circ$ (solid lines); left panel is for models
with absorption only, right panel for absorption + scattering
models.  Dotted lines refer to sandwich models with different
dust-to-starthickness ratios ($\xi$). The straight solid line is for the
screen model.\label{locE}}

\figcaption[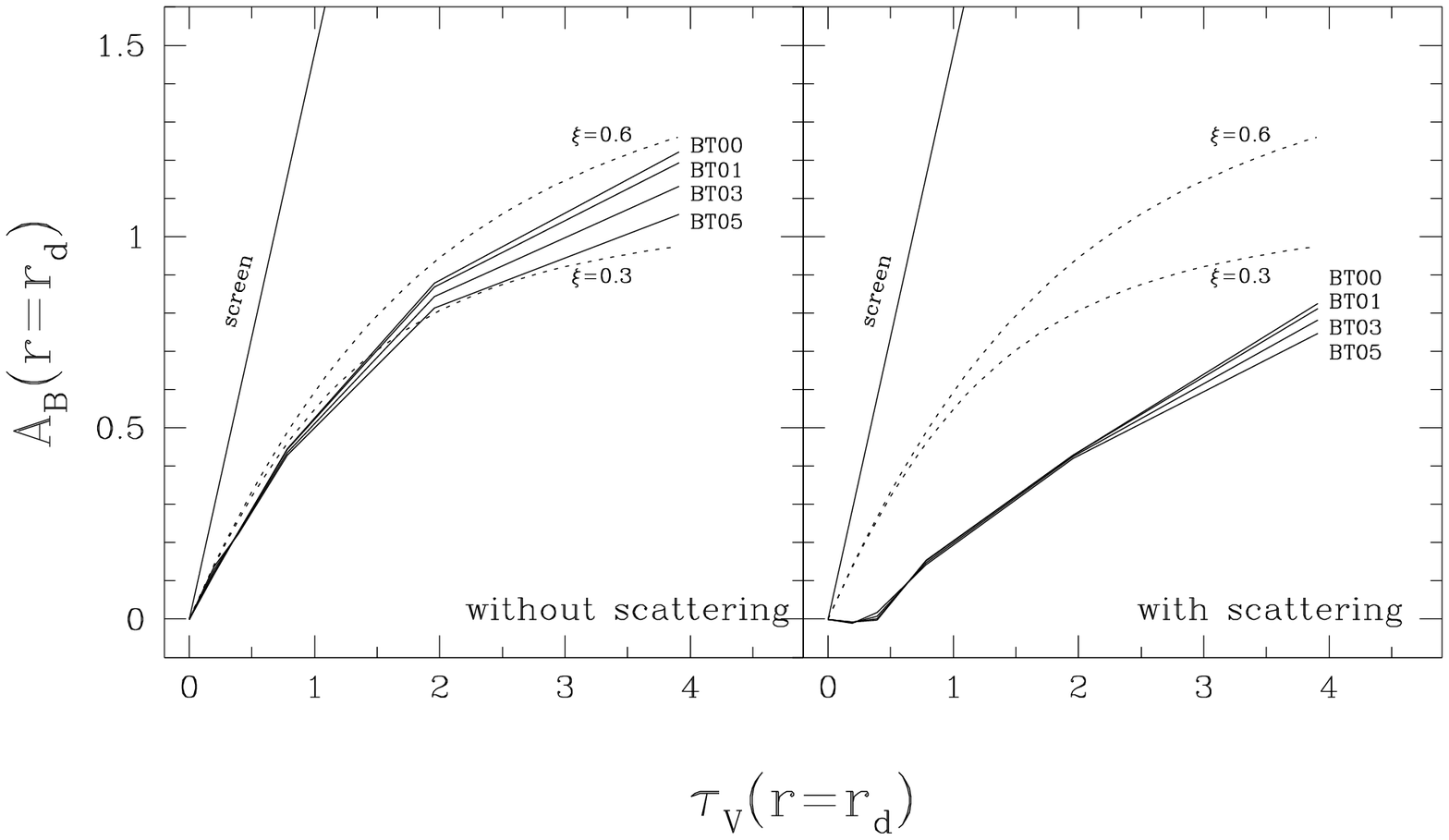]{ As in Fig.~\protect\ref{locE}, but
for $r=r_d$. \label{locE1r0}}

\figcaption[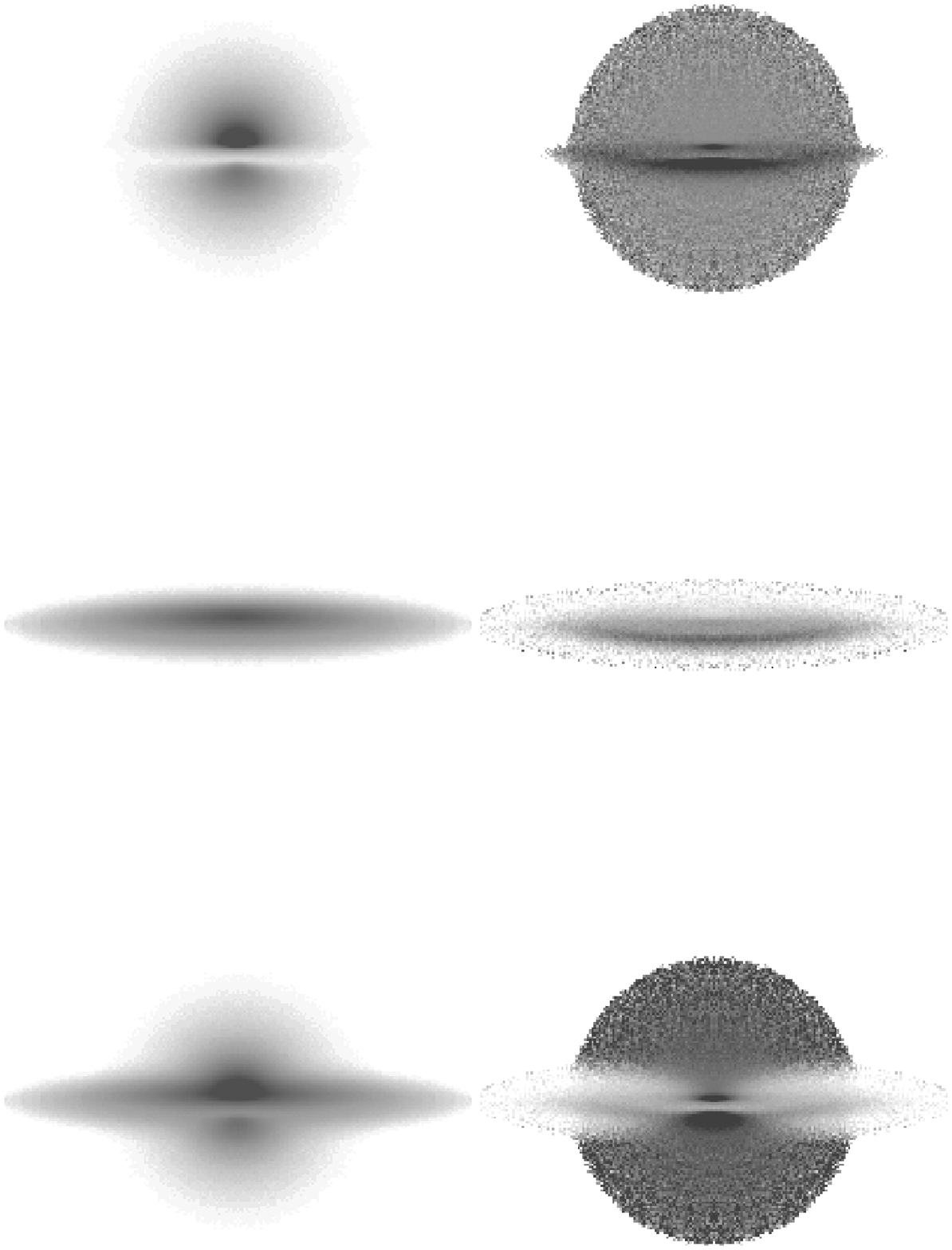]{ B band images (left) and  B-I color maps
(right) for a bulge (top), a disk (center) and  a
BT05 model (bottom) as seen from an inclination of $82^\circ$.
The dust disk is the same in the three cases, with \tc$=10$. \label{compo}}

\figcaption[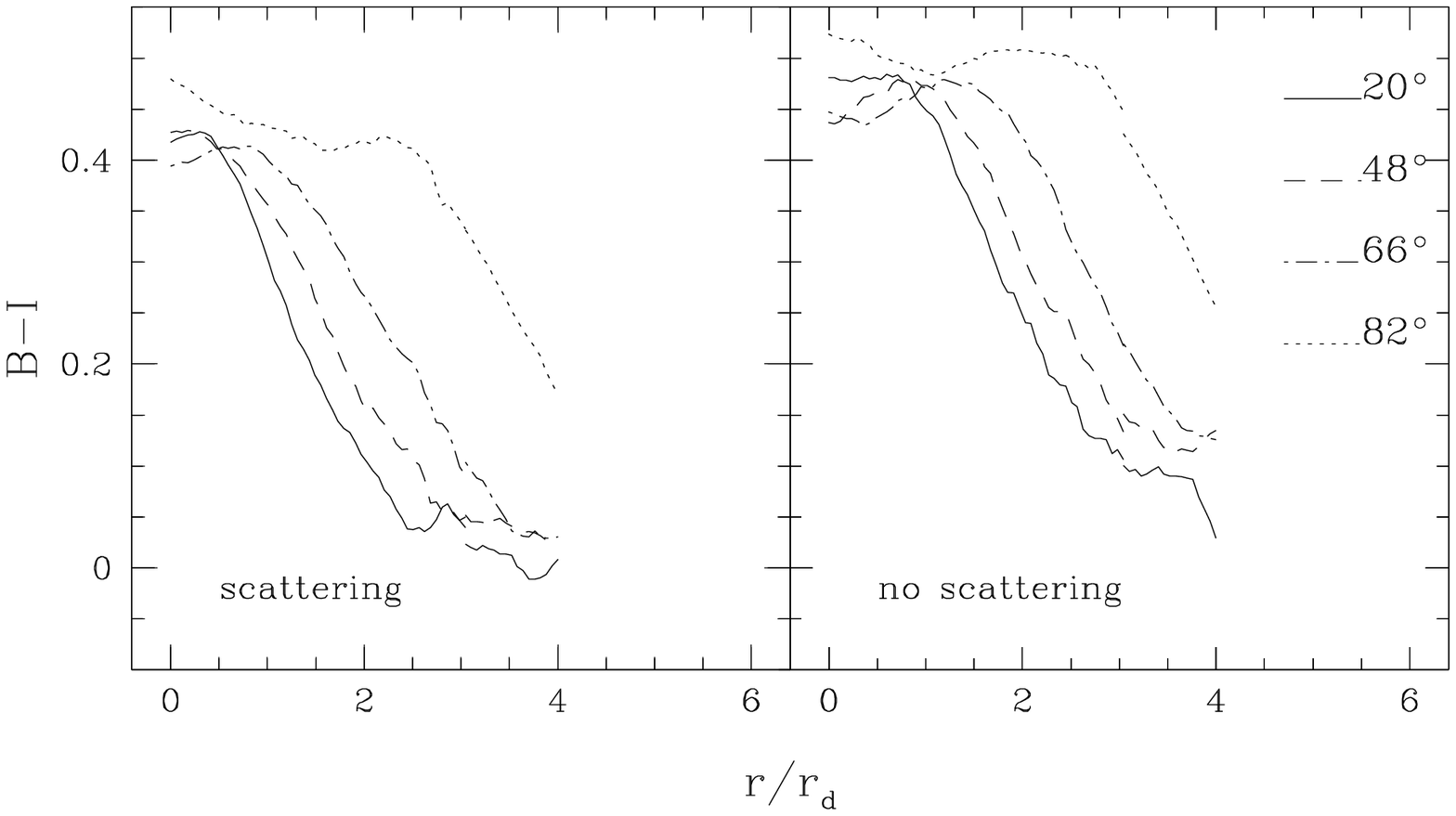]{Color profiles along the major axis for a BT00 model
(pure disk)
with $\tc=5$ at various inclinations. Left panel is for
models with only absorption, the right one for absorption+scattering
model. Lines are truncated at $r= 4r_d$ because of the poor S/N at
larger radii.\label{grad}}

\figcaption[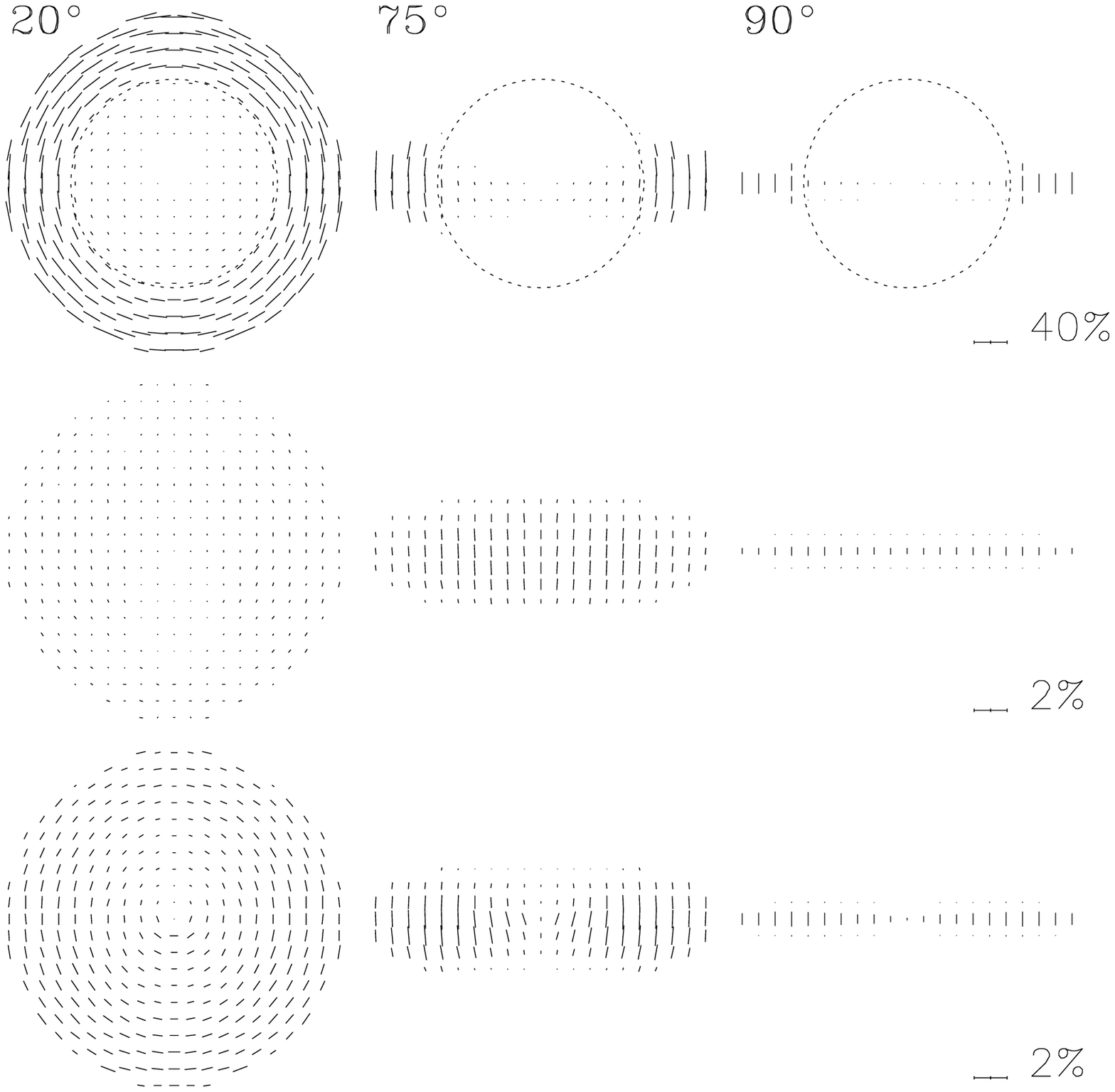]{B-band polarization maps for a bulge (top), a disk
(middle), and a BT05 model (bottom). The optical depth is $\tc=10$ and
the inclinations $20^\circ, 75^\circ, 90^\circ$ (from left to right).
The dust disk is the same  in the three cases.
For each model, the linear polarization scale is given on the right.
The
dashed circle in the upper maps corresponds to the extent of the bulge.
\label{polexa}}

\figcaption[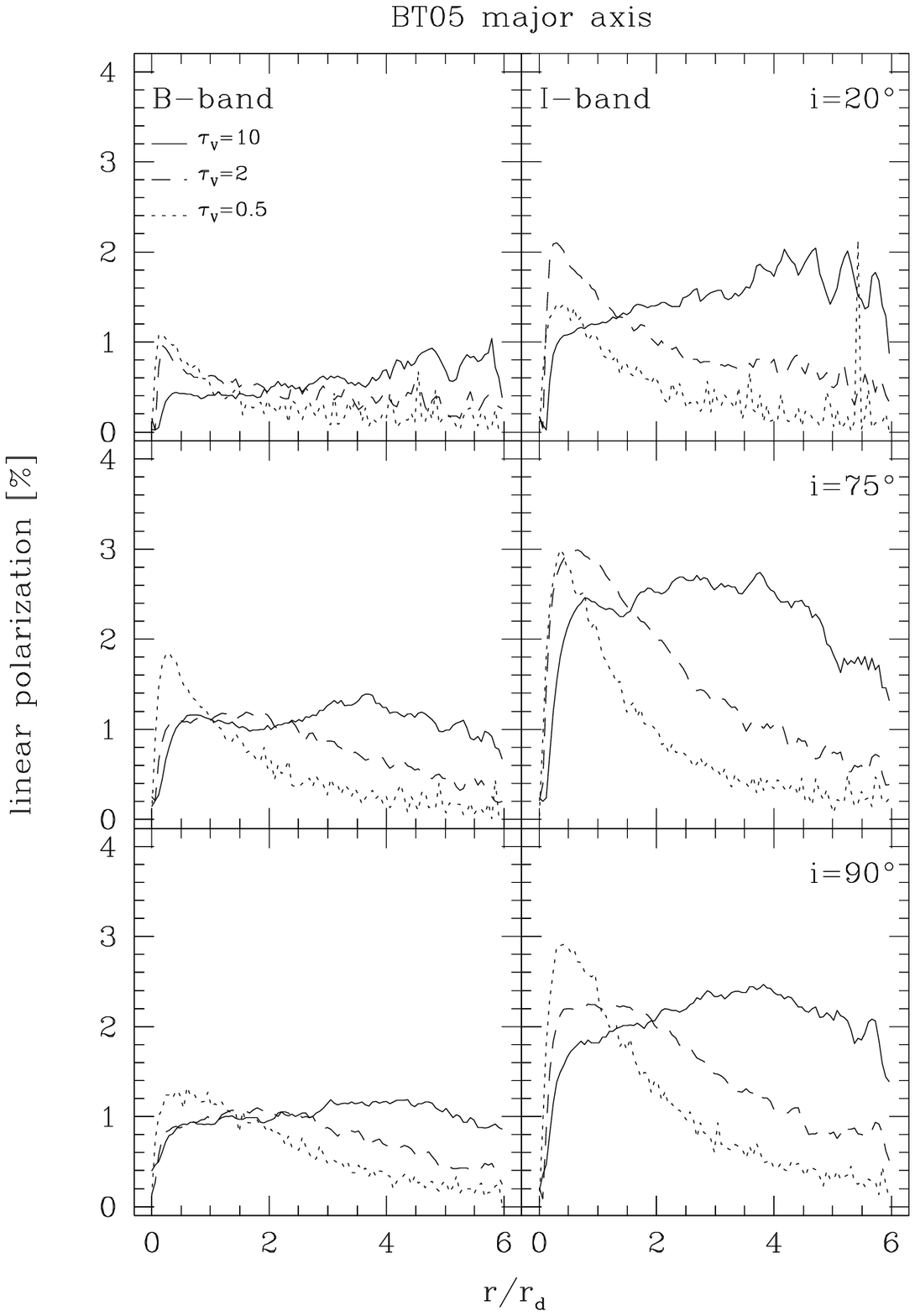]{Linear polarization profiles along the
major axis for BT05
models at inclinations $20^\circ, 75^\circ, 90^\circ$, for B-band (left)
and I-band (right). Profiles for
$\tc=0.5, 2, 10 $ are plotted.\label{polprof}}

\figcaption[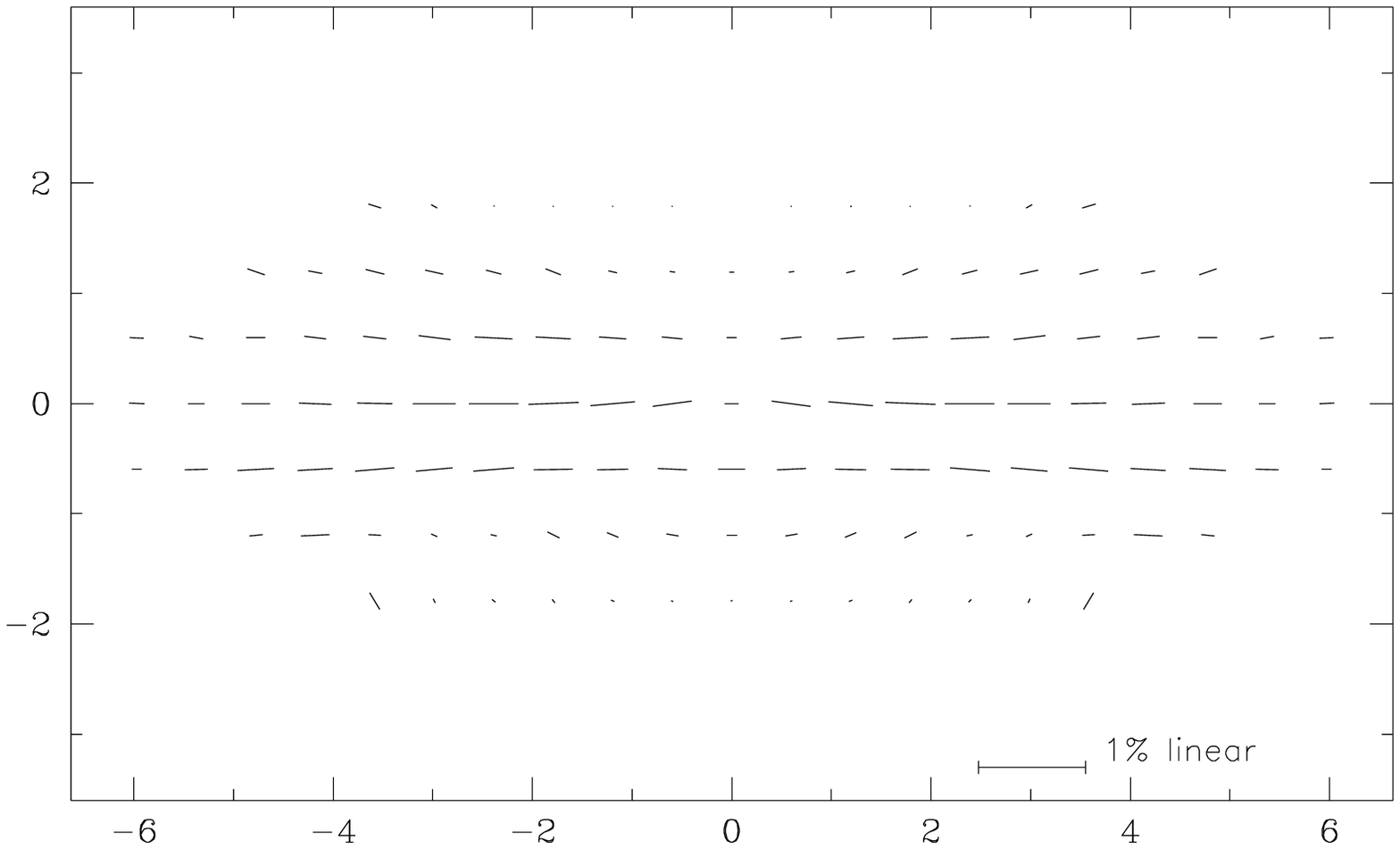]{B-band BT05 model seen from 75$^\circ$, with
$\tau_B(0) = 5$ and a lower grain size cutoff $a_{-}=0.15\mu$m;
coordinates are measured in units of $r_d$.
\label{paralle}}

\figcaption[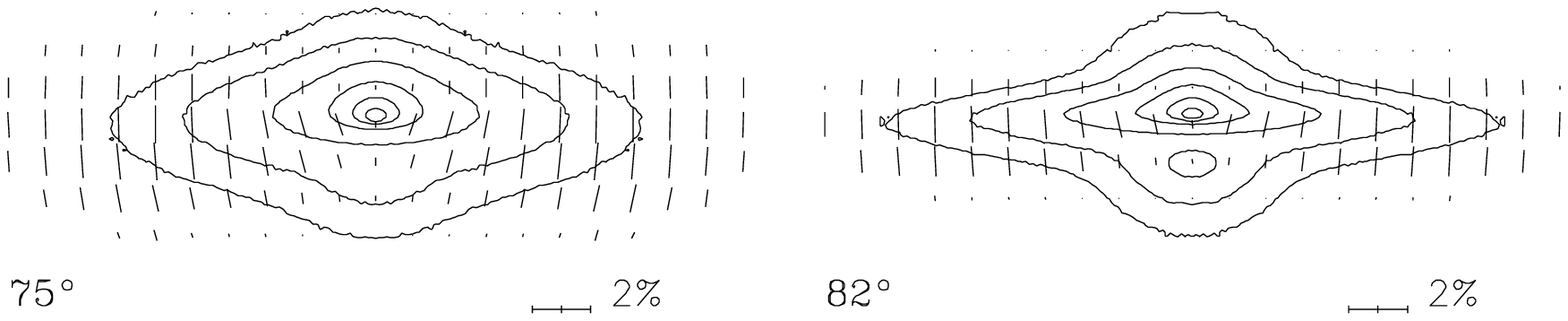]{B-band polarization maps for a BT05 model, with
$\tc=10$, for inclinations $i=75^\circ$ (left) and $i=82^\circ$
(right). Isophotes up to 25 B-mag arcsec$^{-2}$ are superimposed,
spaced by 1 mag.
The assumed face-on disk central brightness is 21.6
B-mag arcsec$^{-2}$ (Freeman 1970 \protect\markcite{fr70}). \label{osserva}}

\figcaption[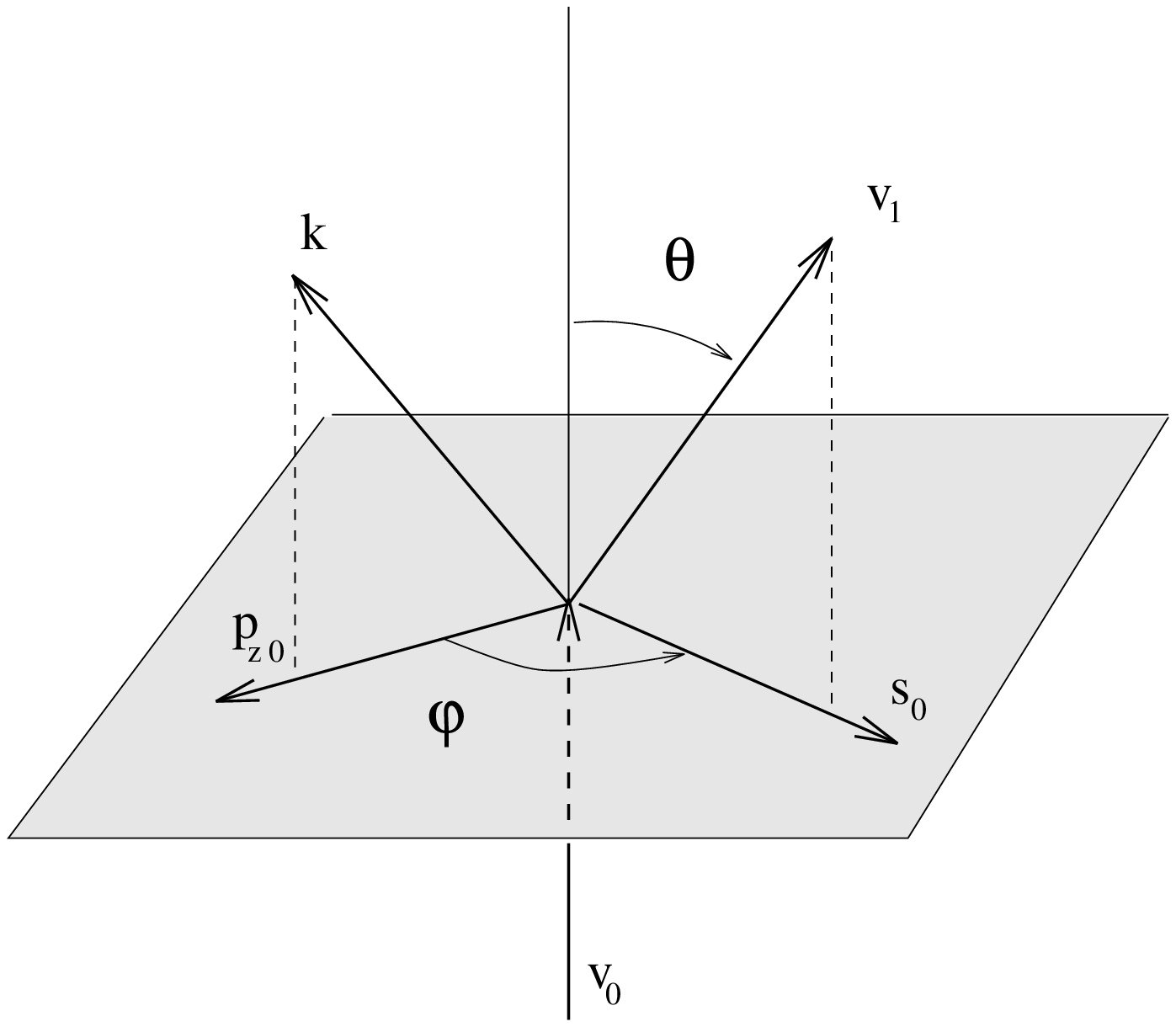]{Scattering geometry: ${\bf v}_0$ is
the original direction of the photon and ${\bf v}_1$ the  direction after
scattering; the other vectors are defined in the text.\label{v0_pz0}}

\figcaption[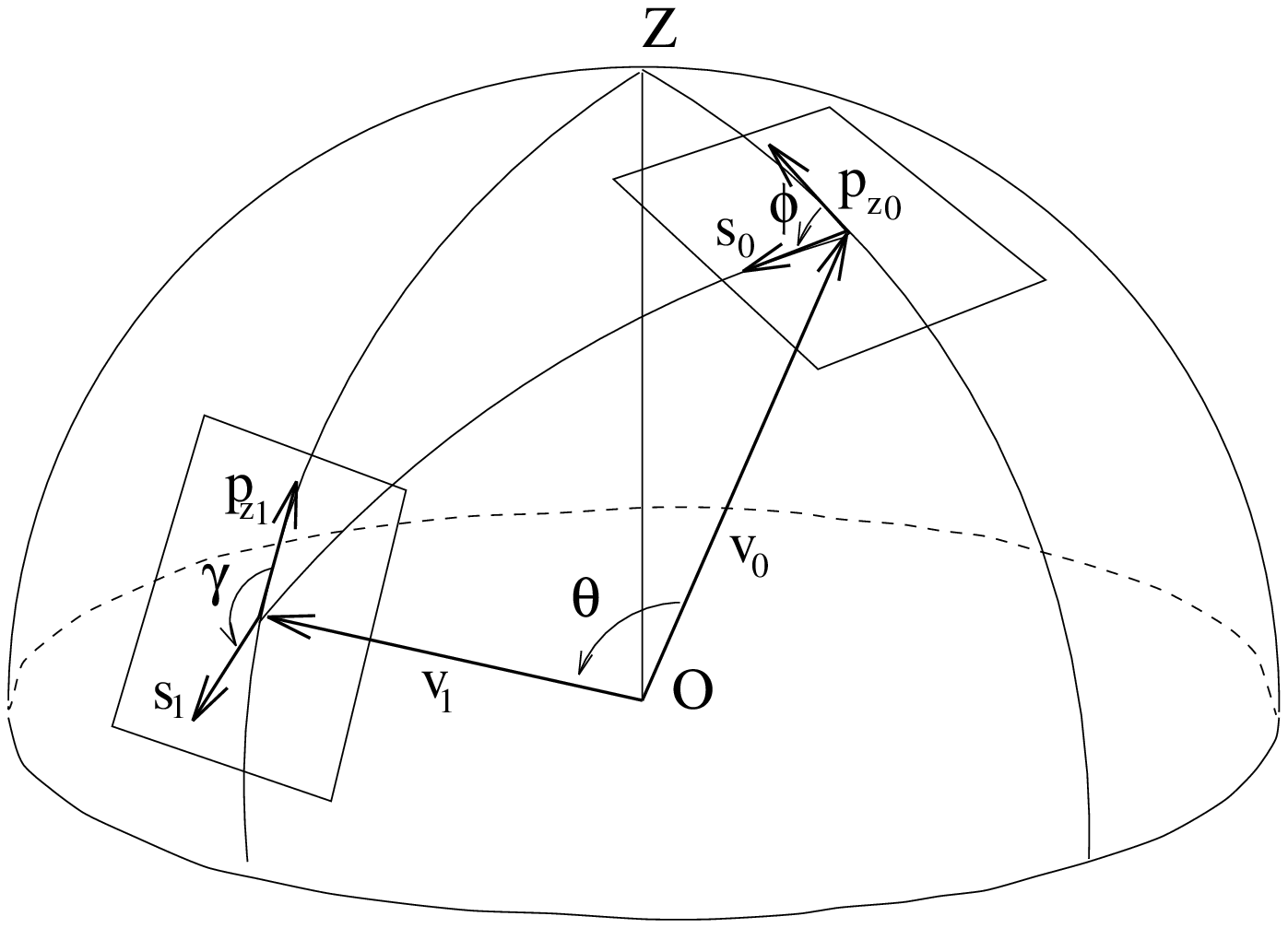]{Scattering geometry in polar representation.
\label{polare}}
\end{document}